\documentclass[12pt,preprint]{aastex}

\begin{document}


\title{The Properties of X-ray Luminous Young Stellar Objects in the NGC 1333 and Serpens Embedded Clusters.}

\shorttitle{X-ray \& IR-Spectral Properties of YSOs in Serpens \& NGC 1333}

\author{E. Winston\altaffilmark{1,2,10}, S. T. Megeath\altaffilmark{3}, S. J. Wolk\altaffilmark{2}, B. Spitzbart\altaffilmark{2}, R. Gutermuth\altaffilmark{4,5},  
L.E. Allen\altaffilmark{6}, J. Hernandez\altaffilmark{7,8}, K. Covey\altaffilmark{2}, J. Muzerolle\altaffilmark{9}, J. L. Hora\altaffilmark{2}, P. C. Myers\altaffilmark{2}, G. G. Fazio\altaffilmark{2}}

\altaffiltext{1}{School of Physics, University of Exeter, Stocker Road, Exeter, EX4 4QL, U.K.}
\email{ewinston@astro.ex.ac.uk}
\altaffiltext{2}{Harvard Smithsonian Center for Astrophysics, 60 Garden St., Cambridge MA 02138, USA.}
\altaffiltext{3}{Ritter Observatory, Dept. of Physics and Astronomy, University of Toledo, 2801 W. Bancroft Ave., Toledo, OH 43606, USA. }
\altaffiltext{4}{Five Colleges Astronomy Department, Smith College, Northampton, MA  01027}
\altaffiltext{5}{Department of Astronomy, University of Massachusetts, Amherst, MA  01003}
\altaffiltext{6}{NOAO, Tucson, AZ, USA.}
\altaffiltext{7}{Centro de Investigaciones de Astronomia, Apdo. Postal 264, Merida 5101-A, Venezuela.  }
\altaffiltext{8}{Department of Astronomy, University of Michigan, Ann Arbor, MI 48109.   }
\altaffiltext{9}{Space Telescope Science Institute, Baltimore, MD, USA. }

\altaffiltext{10}{Visiting Astronomer at the Infrared Telescope Facility, which is operated by the University of Hawaii 
under Co-operative Agreement no. NCC 5-538 with the National Aeronautics and Space Administration, 
Science Mission Directorate, Planetary Astronomy Program.}

\begin{abstract}

We present new {\it Chandra} X-ray data of the NGC 1333 embedded cluster and combine these 
data with existing  {\it Chandra} data, {\it Sptizer} photometry and ground based spectroscopy of 
both the NGC 1333 and Serpens Cloud Core clusters to perform a detailed study of the X-ray 
properties of two of the nearest embedded clusters to the Sun.

We first present new, deeper observations of NGC 1333 with {\it Chandra ACIS-I}  and combine 
these with existing {\it Spitzer} observations of the region.   
In NGC 1333, a total of 95 cluster members are detected in X-rays, of which 54 were previously 
identified in the {\it Spitzer} data.  Of the {\it Spitzer} identified sources, we detected 23\% of the 
Class I protostars, 53\% of the Flat Spectrum sources, 52\% of the Class II, and 50\% of the 
Transition Disk young stellar objects (YSO).   Forty-one Class III members of the cluster are 
identified, bringing the total identified YSO population to 178.    

The X-ray Luminosity Functions (XLFs) of the NGC 1333 and Serpens clusters are compared to 
each other and the Orion Nebula Cluster.  Based on a comparison of the XLFs of the Serpens 
and NGC 1333 clusters to the previously published ONC, we obtain a new distance for the Serpens 
cluster of $360^{+22}_{-13}$~pc.  

Using our previously published spectral types, effective temperatures and bolometric luminosities, 
we analyze the dependence of the X-ray emission on the  measured stellar properties.   The X-ray 
luminosity was found to depend on the calculated bolometric luminosity as in previous studies of 
other clusters.  We examine the dependence of $L_{X}$ on stellar surface area and effective 
temperature, and find that  $L_{X}$ depends primarily on the stellar surface area.   In the NGC 1333 
cluster,  the Class III sources have a somewhat higher X-ray luminosity for a given surface area.  
We also find evidence in NGC 1333 for a jump in the X-ray luminosity between spectral types of M0 
and K7, we speculate that this may result from the presence of radiative zones in the K-stars.   

The gas column density vs. extinction in the NGC 1333 parental molecular cloud was examined 
using the Hydrogen column density determined from the X-ray absorption to the embedded stars 
and the $K$-band extinction measured to those stars.  In NGC 1333, we find  
$N_H$ $= 0.89\pm0.13\times10^{22}$~$A_K$, this is lower than expected of the standard ISM 
but similar to that found previously in the Serpens Cloud Core.

\end{abstract}

\keywords{infrared: stars --- X-rays: stars --- stars: pre-main sequence --- circumstellar matter}

\today

\section{\bf Introduction}

In recent years many studies of young stellar clusters have been undertaken to investigate the emission 
properties of pre-main sequence stars and protostars in the higher energy X-ray regime \citep{wol,get,prefei}.   
These studies compliment those carried out in the mid-IR, where young stars are identified by the excess 
emission produced through the reprocessing of stellar radiation by circumstellar material.  
Optical and near-IR spectroscopic observations of the young stars in these regions add information on the 
fundamental properties of the stars, such as effective temperature and luminosity.  

Young stellar objects (YSOs) often possess  levels of X-ray emission elevated to $ L_{X} \sim 10^4 \times L_{X\odot}$, 
this elevated emission can be used to distinguish  them from field stars  \citep{fei2,fei3}.  
In developed, hydrogen burning stars, X-ray activity arises from magnetic fields generated as a result of shear 
between the core radiative zone and the outer convective zone.  The process behind the generation of the highly 
increased levels of emission in young stellar objects remains uncertain since low mass pre-main sequence stars 
are often fully convective.  Some suggested causes are magnetic disk-locking between the star and disk \citep{hay,iso,rom}, 
accretion onto the star \citep{kas,fav1,fav2}, and alternative dynamo models for coronal emission \citep{kuk,giam}.   

We present a study of X-ray emitting young stars in the Serpens and NGC 1333 clusters.  These two clusters are 
examples of nearby ($<$0.5~kpc), low mass regions with high fractions of protostars yet different spatial 
geometries and frequencies of jets.   Both are deeply embedded in their natal clouds.  
Serpens is the more deeply embedded with extinctions exceeding 40 magnitudes in the visual.   IR and submm 
observations identify at least 38 protostars in the central core region, and more than 130 embedded young 
stars in total \citep{tes1,tes2,dav1,hog,winston}.  
An age of $\le$2~Myrs has been found for the cluster \citep{kaa,win08a}, with some evidence for a halo of older 
sources surrounding the central core \citep[see][for a discussion of the possible age spread.]{win08a} 
The Chandra X-ray observations of Serpens have previously been reported by \citet{gia} and \citet{winston}.  
NGC 1333 contains a high number of embedded protostars, cores and outflows \citep{knee}, and while showing a 
spread in isochronal ages was found to have a median age of 2~Myrs \citep{win08a}.  
The region has previously been observed in X-rays with ROSAT by \citet{prei97} who identify 16 YSOs.   
A study by \citet{get2} with {\it Chandra} associated 95 detections with cluster members and found no difference in X-ray 
luminosity between Classical T-Tauri stars (CTTS) and weak line T-Tauri stars (WTTS).   \citet{pre03} observed NGC 1333 
using XMM, detecting 86 sources in the region.   Optical and IR surveys to identify the YSOs in the region have been 
carried out by \citet{aspin},  \citet{wilk2},  \citet{lada},  \citet{gut07} and \citet{gut09}.   
Over 160 young stars have been identified in the cluster.   NGC 1333 does not appear to be centrally condensed though 
the protostars are observed to trace the underlying gas distribution, whereas the Serpens protostars are located along a 
filament in the centre of the cluster.   

There are four main results in this paper.  First, we utilise the elevated X-ray luminosity to identify YSOs in the 
NGC 1333 region that do not exhibit IR-emission from a dusty disk (evolutionary Class III) and would otherwise 
be indistinguishable from field stars using {\it Spitzer} photometry. Protostars (Class 0/I and Flat Spectrum) and 
pre-main sequence stars with disks (Class II and Transition Disks) may also show elevated X-ray emission and 
we identify such objects with detectable X-ray emission.

Second, we use the X-ray data to redetermine the distance to the Serpens cluster.  One of the 
more elusive properties of Young Stellar Clusters (YSCs) is an accurate measure of their distances.    
The distance underpins luminosity calculations, and therefore the determination of the stellar isochronal 
ages and masses.   The distance to NGC 1333 has recently been measured based on a VERA measurement 
of maser parallax toward SVS 13 to be at 240~pc, and this is the distance used in this study  \citep{hirota}.     
A method for determining the distance to YSCs based on the comparison of the X-ray luminosity 
function (XLF) to a 'universal' XLF \citep{fei} is applied to the Serpens data to obtain a more accurate distance to the cluster.  
The most commonly adopted distance to the Serpens Cloud is 260~pc  based on an extinction versus distance diagram 
to $\sim$450 stars using Vilnius seven-color photometry \citep[a discussion of the distance is given in][]{strai}.   
Other estimates have ranged from 700~pc \citep{zha} to 440~pc \citep{racine,strom} and 310~pc \citep{del}.  
Here we will attempt to judge the distance via a technique applied to the YSOs in the cluster themselves.  

In the third section, we present a detailed study of the combined spectral and X-ray properties of the populations 
of YSOs as a function of their evolutionary class in the Serpens and NGC 1333 clusters, identified by observations 
with {\it Spitzer} and {\it Chandra} and published by \citet{gut08} and \citet{winston,win08a}.  {\it Chandra} has 
recently examined the X-ray properties of the Orion Nebula Cluster (ONC), in a project known as the {\it Chandra} 
Orion Ultra-Deep Project (COUP).   Although the Serpens and NGC 1333 samples are much smaller 
than the COUP survey, the analysis of these two smaller clusters has the advantage of excellent 3.6-24~${\mu}m$ 
photometry; observations longward of 4.5~${\mu}m$ are difficult in the centre of the ONC due to the bright infrared 
emission from the molecular cloud. This photometry allows us to accurately ascertain the evolutionary class of each 
object, and study the dependence of the X-ray properties on evolutionary state. Further, optical and infrared
spectroscopy has been obtained and used to determine the effective temperatures, bolometric luminosities, and 
stellar surface areas of a number of the YSOs. With these data, we can examine the X-ray flux to bolometric 
luminosity by evolutionary class, and examine the dependance on surface area and effective temperature in these 
two clusters.

Finally, we derive the relationship between Hydrogen column density derived from X-ray data and near-infrared extinction.
Consistent with our preliminary work, we find that in these molecular clouds, the hydrogen column density
per magnitude of K-band extinction is $\sim$1/3  that in the diffuse ISM.  This may be the result of grain
growth and coagulation in cold, dense molecular clouds.

\section{\bf Observations and Data Reduction}

\subsection{ {\it Spitzer} \& 2MASS IR Photometry}

We have obtained {\it Spitzer} images of the Serpens and NGC 1333 regions in six
wavelength bands: the 3.6, 4.5, 5.8 and 8.0~$\mu$m bands of the
Infrared Array Camera (IRAC; \citet{faz}) and the 24~$\mu$m and
70~$\mu$m bands of the Multi-band Imaging Photometer for {\it Spitzer} (MIPS; 
\citet{rie}).  The photometry extracted from these data was
supplemented by $J$, $H$ and $K_S$-band photometry from the 2MASS point
source catalogue \citep{skr}, resulting in data in nine photometric
bands spanning 1-70~$\mu$m.  The observations, image reduction and photometry 
for the {\it Spitzer} data have previously been presented by \citet{winston} and \citet{gut07}. 
In Serpens, \citet{winston} identified 117 YSOs with IR-excesses: 22 Class 0/I objects, 16 
Flat Spectrum, 62 Class II and 17 Transition Disks \citep[see][on contamination of Transition 
Disk objects by AGB stars.]{win08a}  
In comparison, \citet{gut07} identified 137 YSOs in NGC 1333: 11 candidate Class 0, 26 Class I, 
16 Flat Spectrum, 80 Class II and 4 Transition Disks \citep[note that the Flat Spectrum classification 
was applied to the NGC 1333 catalog by][]{win08a}.

\subsection{ {\it Chandra} Data Reduction}

The X-ray data were taken from the $Chandra$ ANCHORS (AN archive of CHandra Observations of Regions 
of Star formation\footnote{http://cxc.harvard.edu/ANCHORS/}) archive. 
The Serpens cluster was observed on June 19th 2004, OBSID 4479, with a 88.45~ks exposure time,  
the field centered on the $J2000$ coordinates:  $18^{h}29^{m}50^{s}, +01^{d}15^{m}30^{s}$.  
The Serpens observations have been previously published in \citet{gia} and \citet{winston}. 
The NGC 1333 data is newly published here, hence we describe the reduction of the NGC 1333 
data in detail.    
NGC 1333 was observed in three epochs: OBSID 6436, a 36.95~ks exposure on July 5th 2006,  OBSID 6437, 
a 40.12~ks exposure on July 11th, 2006, and OBSID 642, a 43.91~ks exposure on July 12th 2000, giving a total 
exposure time of 120.98~ks.   The 2006 fields were centred on the $J2000$ coordinates: $03^{h}29^{m}02.0^{s}, 
+31^{d}20^{m}54^{s}$, with the 2000 field centred on the $J2000$ coordinates: $03^{h}29^{m}05.6^{s}, 
+31^{d}19^{m}19^{s}$.   These three epochs are combined to provide the final source list used in this analysis, 
with a total of 180 point sources detected in the field.   
Table \ref{tablenp} lists the identifiers, coordinates and properties of the 180 sources identified in NGC 1333.  
The table provides the {\it Chandra} identifier, source locations, the raw and net number counts (net counts are 
background subtracted and aperture corrected), plasma temperature ($kT$), hydrogen column density ($N_H$),  
absorbed and unabsorbed X-ray flux ($F_X$), and flaring statistics.   The field of view of {\it Chandra} is 17$'$, smaller 
than that covered by the IR data: thus not all IR-identified cluster members were observed at X-ray wavelengths.

\subsubsection{Source Detection}
For NGC 1333, the three separate epochs were first combined using the merge\_all script\footnote{Chandra contributed software - http://cxc.harvard.edu/ciao/download/contrib.html}
This wrapper script uses standard {\it ciao} tools to create a merged event file.  The inputs were the individual 
event files and the level 1 aspect solution data products.  The event files were standard level 2 products with
an energy filter of 0.3-8.0 keV and a spatial filter to select CCD chips 0-3 applied.  We used obsid 6437 
(the latest observation) as the coordinate reference to which the other observations have their coordinates 
reprojected.  With the merged event file we ran a recursive blocking source detection algorithm.  In this scheme,  
we performed two iterations using the ciao wavlet based source detection tool, {\it wavdetect}.  The first pass looked 
at a square region 15' on each side and centered at the aimpoint with full resolution.  
The second pass examined a $25' \times 25'$ area including all six ACIS chips with pixels binned by two.  

Source and background regions were determined separately for the three datasets. The program $dmcoords$ 
was used to convert the WCS coordinates of the merged events list back to the physical coordinates in the 
individual observations. 
From the detected source positions, we create elliptical source extraction regions for the 3 individual 
datasets.  For each source, the parameters of the ellipse are the size of the semimajor and semiminor 
axis as well as the orientation angle of the ellipse, $\rho$.  \citet{allen} have produced a parameterization 
of the axial description of the Chandra point spread function by using observed data. The ellipse 
semimajor and semiminor axes for a given encircled energy fraction (ECF) and energy can be 
interpolated from the lookup table as a function of azimuth angle, $\phi$, measured in degrees, and 
off-axis angle, $\theta$.\footnote{A lookup table of the parameterization is available at: http://cxc.harvard.edu/cal/Hrma/psf/ECF/hrmaD1996-12-20hrci\_ell\_ecf\_N0003.fits}
The size of the ellipse axis are determined by a linear interpolation.  The table value for the off-axis 
angle nearest the source, at 95\% ECF and 1.0 keV is used.  The interpolation is then between the 
nearest adjacent azimuth angles.
Due to the coarse grid of this calibration parameterization, it is impractical to interpolate the proper 
ellipse angle, $\rho$.  Therefore, we used the CIAO tool $mkpsf$ to obtain images of the PSF at a finer 
grid of off-axis and azimuthal angles around the ACIS array.  We fit ellipses at each location, and derived 
a general empirical formula:  $\rho = (\phi \times 0.47 + 146.6 - ROLL + 360) mod 180$.
A centroid algorithm is then applied to the source ellipse to balance the distribution of photons within the 
ellipse in the actual dataset.
Background regions were defined as similar elliptical annuli, centered on the source position.  The outer 
semimajor and semiminor axes were 6 times the size of the source ellipse and the inner axes 3 times the source. 
Finally a check was done to identify nearby or overlapping sources.  These sources are removed from the source 
extraction area with exclusion regions of ellipses 3 times the original source size.
Finally, {\it psextract} was run to create appropriate files for spectral fitting. 
{\it Psextract} follows the standard CIAO 3.4 threads\footnote{http://cxc.harvard.edu/ciao3.4/threads/ispec.html}. 
First,  {\it dmextract} is used to create the source and background spectra in Pulse Height Amplitude (PHA) space.  
Then, {\it dmstat} lists the chip location of the source and background regions and  {\it acis\_fef\_lookup} uses this 
information to identify the appropriate spectral calibration files (FEFs).
Then {\it mkrmf} applies the FEF to create the source and background Response Matrix Function (RMF).  
These files are used by {\it Sherpa} to convert PHA channels to energy.  In most cases, because the source and 
background regions have the same average detector coordinates, the two files are the same.  Next, {\it asphist}  
is used to create the aspect histogram, which is a binned representation of aspect motion during the observation. 
This is then applied by {\it mkarf} to create an appropriate Ancillary Response Function (ARF).
The ARF  contains the combined telescope/filter/detector areas (``effective area'') and the quantum efficiency (QE) 
as a function of energy averaged over time.
  
\subsubsection{Spectral Analysis}

We used the CXC SHERPA program to generate spectral fits and solve simultaneously for the absorbing column 
of hydrogen, $N_H$, and the temperature of the thermal plasma, $kT$.   The results incorporate simultaneous fits 
of the three observations by defining each spectrum independently and fitting with the same model for a result which  
minimizes the combined residuals.   We fit the unbinned spectra to a one temperature Raymond-Smith model.   
In the case of NGC 1333, the temporal spread in the data means that the temperature and flux of the sources could 
change between epochs;  the reported fits are effectively an average of the data from the epochs.   
We used unbinned data fitted using CSTAT statistics and Powell optimization.  
This combination is more robust in avoiding local minimizations, which  binned methods tend toward.  This also gives 
more reliable results in low count  sources.  This later point was especially important for NGC 1333 in which the 
observations were thrice divided and hence we would have required about 30 counts in each observation for 
successful binned fits.    Recent work by  Siemiginowska (2008; http://cxc.harvard.edu/sherpa/references/papers/statistics.pdf)
also indicates that unbinned statistics are less biased then the binned form.   We used a model background consisting 
of a constant, power-law, and Gaussian to resemble a typical ACIS background.   X-ray fluxes are estimated using 
{\it Sherpa's} {\it eflux} function.   We calculate the total integrated energy flux over our standard energy range (0.3-8.0 keV).  
A flux is returned in units of $ergs/cm^2/sec$.  This  is repeated using only the kT model component to obtain the 
unabsorbed flux.  The values are scaled up from the 95\% encircled energy radius for the final published values.

\subsection{Optical \& Near-IR Spectroscopy}

To undertake a more thorough examination of the X-ray properties of the two young
clusters, we obtained spectral observations of members identified by the {\it Spitzer} and {\it Chandra}
observations (as well as a number of stars surrounding the clusters to search for additional
members). The spectral observations of both clusters were obtained using the Hectospec instrument
on MMT and the SpeX instrument on IRTF. A full analysis of these data can be found in the companion
paper \citet{win08a}, where a comparison with the IR photometry and evolutionary
classifications of the sources is undertaken. Hectospec, on the MMT, provides simultaneous
far-red (6000-9000 \AA) spectra of up to 300 objects, and was used to observe the less embedded 
known cluster members and to search for diskless members not detected by Chandra.  
SpeX on the IRTF provided spectra of known, more deeply embedded
Class II and III members, over the $H$, and $K$-bands. A brief summary of the relevant points
from the companion paper is provided: In total,  spectral types  were obtained for 64 young stellar objects (YSOs) 
in Serpens, with a further 74 in NGC 1333. All of the SpeX sources were known cluster members, 
while eight (4 in Serpens, 4 in NGC 1333) of the Hectospec sources are possible new
diskless young stellar objects that were not detected in the {\it Chandra} data but show $Li~I$
absorption, an indicator of stellar youth. 
We determined spectral types for 60 and 70 previously identified YSOs in Serpens
and NGC1333, respectively.
Of these members with spectral types,
34 and 59 have {\it Chandra}  X-ray counterparts in Serpens and NGC 1333, respectively.

\section{\bf X-ray Characteristics of the NGC 1333 YSOs \label{xrayc}}

\subsection{YSO Detection Rates}

Previous infrared studies by \citet{gut07} have revealed 137 YSOs with IR-excess emission 
in the NGC 1333 cluster.  
The {\it Chandra} observations of the region identified 180 point sources, of which 95 were matched 
to an IRAC or 2MASS detection in at least one IR band, with a maximum separation between the 
{\it Chandra} and IR positions of  $1.12''$.  
Table \ref{tablenp} lists the coordinates and X-ray properties of the 180 X-ray point sources in the 
{\it Chandra} catalogue.  Table \ref{tablenpi} lists the {\it Chandra} identifier and coordinates, the 
infrared coordinates and IR photometry of the 95 X-ray detections with an IR counterpart.  
The formal uncertainties for each {\it Spitzer} magnitude are listed; in addition to these uncertainties 
is an approximately 5\% uncertainty in the calibration.   
The remaining 85 X-ray sources were not detected in any IR band and were randomly distributed 
across the field of view, and are therefore considered to be background contamination from AGN.  
In this paper we use the classification scheme used in \citet{winston}: Class 0/I sources, with rising 
spectral energy distributions (SEDs) with $\alpha > 0.3$ where $\alpha = \frac{d\lambda F_{\lambda}}{d\lambda}$,     
Flat Spectrum sources with flat SEDs and $-0.3 < \alpha < 0.3$, Class II sources with decreasing SEDs ($\alpha < -0.3$), 
Transition Disks which show weak infrared excesses at 3-8~$\mu$m and are thought to have large holes in their 
inner disks,  and Class III members of the cluster which lack an infrared excess.   

Of the remaining 95 X-ray sources with an IR counterpart, 54 were known IR YSOs.  
The remaining 41 sources exhibited no IR-excess emission and were identified as YSOs solely by their 
elevated X-ray emission.  These sources are Class III diskless members of the cluster,  bringing the 
total number of cluster members to 178.  
Of the YSO detections, 64 were previously identified in the \citet{get2} study, including 22 of the Class III 
young stars.  Thus we identify 19 new Class III members.      
Of the cluster members, $41/178$ or 23\% were identified solely by Chandra. 
A list of the {\it Spitzer} coordinates and identifiers of the YSOs detected in the {\it Chandra} field are 
given in Table~\ref{tableids} along with the evolutionary class and \citet{get2} identifier.  

The {\it Chandra} ACIS-I FOV did not cover the entire {\it Spitzer} IRAC field.  
Of the 137 IR-excess YSOs, 127 were in the ACIS FOV:    
None of the eleven cluster members determined to be in evolutionary  Class 0 
were detected.  Five of the twenty-two Class I protostars were detected (22.7~$\pm$~10\% of the 
Class I's in X-ray field of view);  eight of the fifteen Flat Spectrum sources were observed  (53.3~$\pm$~19\%). 
Thirty-nine of the seventy-five Class II objects were found to have X-ray counterparts (52\%~$\pm$~9\%), 
while two of the four Transition Disk objects were detected (50~$\pm$~35\%).  
These are similar detection fractions to the $\sim$50\% found in Serpens \citep{winston}. 
No difference in detection rate was observed between the later evolutionary classes; to within 
1~$\sigma$, approximately 50\% of objects in each of these classes were detected. The X-ray detection 
fraction in NGC 1333 is slightly lower in the Class I group; however, the protostars in NGC 1333 are more dispersed 
than in Serpens and tend to lie further off-axis in the ACIS field of view.  
Compared to the later evolutionary stages, the protostars are more deeply embedded and their soft X-ray 
emission is more highly absorbed. Further, it is possible that the NGC 1333 protostars are at an early evolutionary stage, the 
mechanism of X-ray production in these stars may just be turning on in this cluster \citep{gia}.

\subsection{Spatial Distribution}

\begin{figure}
\epsscale{1.}
\plotone{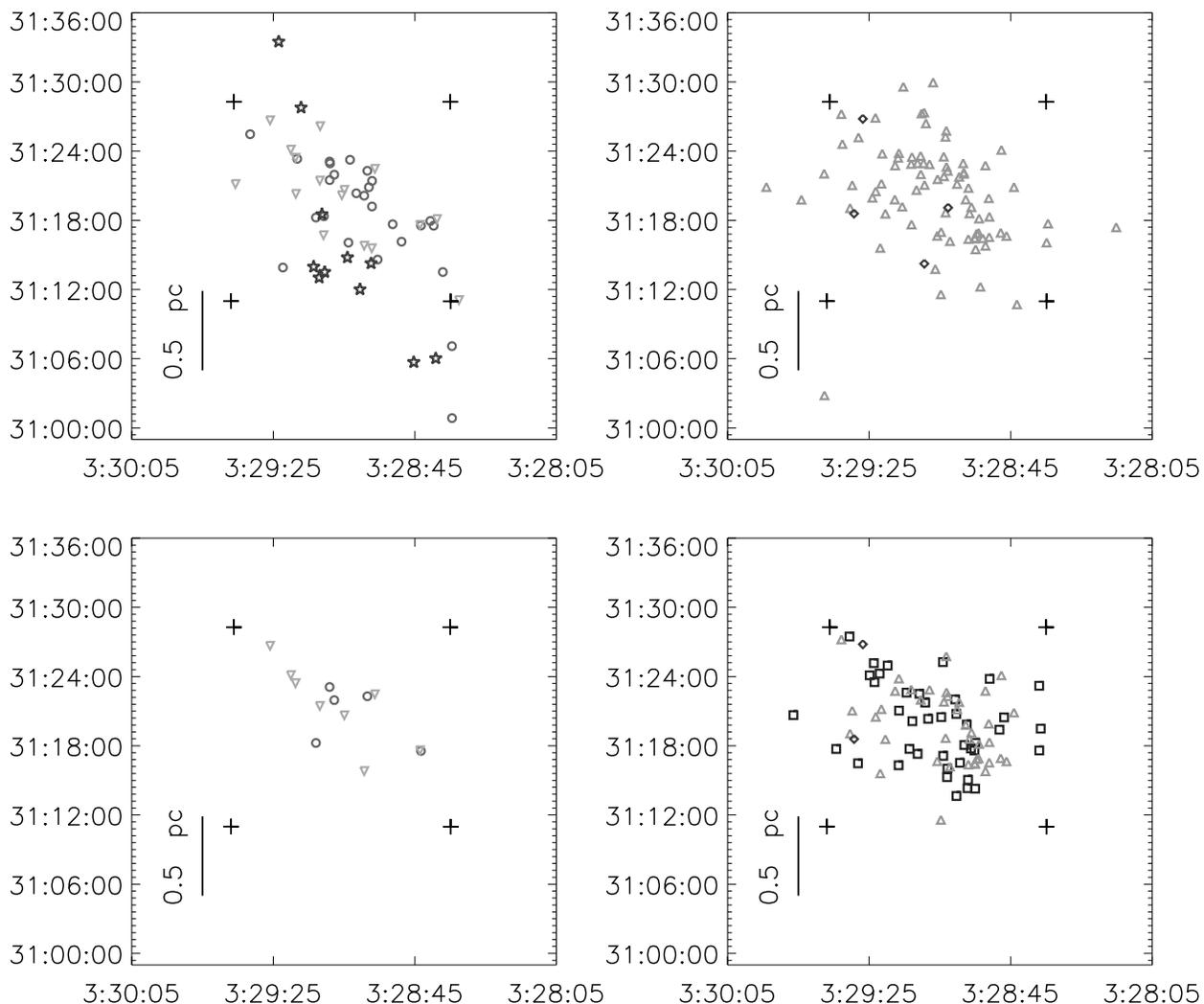}
\caption[Spatial Distribution of YSOs in NGC 1333]{ {\it Above}: Graphs of the spatial distribution 
  of the various classes of YSOs in NGC 1333. {\it Upper Left}: Class 0, Class I and Flat Spectrum sources, in 
  stars, circles and inverted triangles, respectively. {\it Upper Right}: Class II (triangles) and Transition 
  Disk members (diamonds). {\it Lower Left}: X-ray selected sample of Class 0/I and Flat Spectrum 
  objects. {\it Lower Right}: X-ray selected sample of the Class II, Transition Disk, and Class III 
  members (shown by squares).  The elongated distribution of the protostellar objects can be 
  observed, while the Class II and later classes follow a more dispersed pattern over the region. 
  The four crosses outline the {\it Chandra} FOV.  }
\label{figsdn}
\end{figure}

The 178 young stellar objects identified as belonging to the NGC 1333 cluster were 
mapped to determine their spatial distribution, as shown in Fig.~\ref{figsdn}. 
The spatial distribution is shown for the sample of infrared-excess stars (upper two plots) and 
the sub-sample of X-ray luminous member stars (lower two plots), including the Class III objects.  
The {\it Chandra} field of view is constrained to 17$'$$\times$17$'$, a region outlined by the four crosses 
in the figure.    A detailed discussion of the underlying distribution of the NGC 
1333 YSOs is presented in \citet{gut07,gut09}.   
The protostars in NGC 1333 follow an elongated distribution, tracing the underlying filamentary dust and gas 
distribution \citep{hatchell}.  The Class II and the newly identified Class III stars are more concentrated into a 
central 'double' cluster \citep{gut07,lada}.  

\begin{figure}
\epsscale{0.5}
\plotone{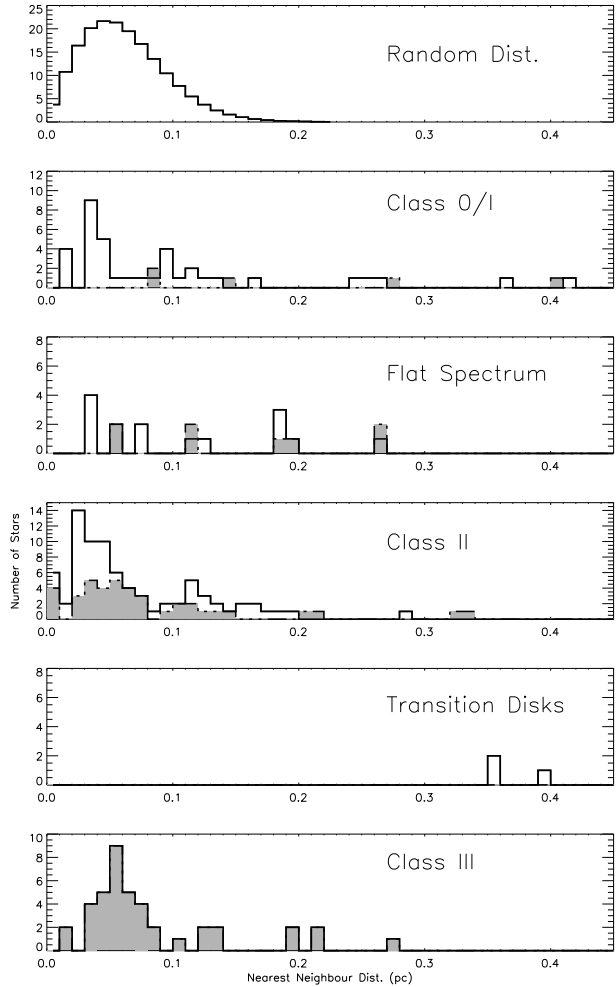}
\caption[Nearest Neighbour Distribution of YSOs in NGC 1333]{ Histograms of the nearest neighbour 
distances of the NGC 1333 YSOs by advancing evolutionary class: Class 0/I, Flat Spectrum, Class II, 
Transition Disk, Class III. The topmost graph plots the random distribution. The solid grey histograms 
indicates the distribution of the nearest neighbor distances calculated using only the X-ray detected 
sources in each class. }
\label{figsdc}
\end{figure}

The spatial distribution of the cluster sources was examined for each of the evolutionary
classes using a nearest neighbour technique. In keeping with our Serpens study \citep{winston},
the Class 0 and Class I protostars were combined into one group.  The nearest neighbour 
distance is the projected distance to the nearest YSO of the same evolutionary class, using the 
adopted distance to NGC 1333 of 240~pc.   Figure \ref{figsdc} shows the distribution of nearest 
neighbour distances.  We also calculate the nearest neighbor distance for the sub-sample of 
X-ray detected sources in each evolutionary class.  
The mean separation between the combined group of Class 0 
and Class I protostars is 0.099~pc, with the separation between the Flat Spectrum objects at 
0.139~pc.  In Serpens, the equivalent separations were 0.033~pc and 0.109~pc, respectively  (the 
Serpens separations are the results from \citet{winston} scaled up to account for the revised 
distance of 360~pc to Serpens, c.f. Section~\ref{xlfdist}).  
The Class II sources have a slightly smaller average separation than the protostars of only 0.08~pc 
(in Serpens 0.134~pc).   
The separations between the Transition Disk stars have a mean value of 0.437~pc (this is biased 
by the small number  of sources), in Serpens it was 0.183~pc.  
The mean separation for the Class III stars is 0.097~pc, similar to that of the youngest protostars.  
In Serpens, the Class III stars had a mean separation of 0.181~pc, five times that of the protostars.   
The mean separation of the Class III sources may be biased to a lower value by the field of view
of the Chandra data. The X-ray field covers 90\% of the Class II sources however, so unless the 
Class III distribution is greatly extended beyond that of the Class II, then this value is likely a 
representative result. 

Comparison of the spatial distributions of the YSOs in each evolutionary class with a random distribution 
was carried out by generating 10,000 random distributions with an equal number of stars to that of the 
given evolutionary class.  For the Class III sources the size of the random field is constrained to the 
{\it Chandra} field of view.  The resulting nearest neighbour distributions were compared using the K-S test 
and the mean probability taken.  The Class I, Class II, and Class III populations were all unlikely to arise 
from a random distribution with probabilities $<  5 \times 10^{-4}$.  The Flat Spectrum and Transition Disk 
sources had probabilities of 24\% and 53\%, due to the small number of sources in each class.

In distinct contrast to Serpens, the  NGC 1333 star forming region shows a  cluster arrangement 
where the young protostars show a similar spacing to that of the Class II  \& III populations.
It is important to note that the Class II and Class III sources show very similar spacings.  
We have thus extended the result of \citet{gut08}, that the protostars and Class II objects show 
similar spacings, to the Class III objects.   
The Kolmorogov-Smirnov probability that the Class II and Class III distributions arise from the same 
parent distribution is 26.5\%.    \citet{win08a} also find similar age distributions for the Class II and III objects.  
These observations, taken in aggregate, show that the populations of pre-main sequence stars with 
and without disks exhibit indistinguishable spatial and age distributions; this suggests that the Class III 
objects are not typically older, more evolved stars, but stars that lost their optically thick disks more 
rapidly.

\subsection{X-ray Absorption, $F_X$, and $kT$ of Bright X-ray Sources}

\begin{figure}
\epsscale{1.}
\plotone{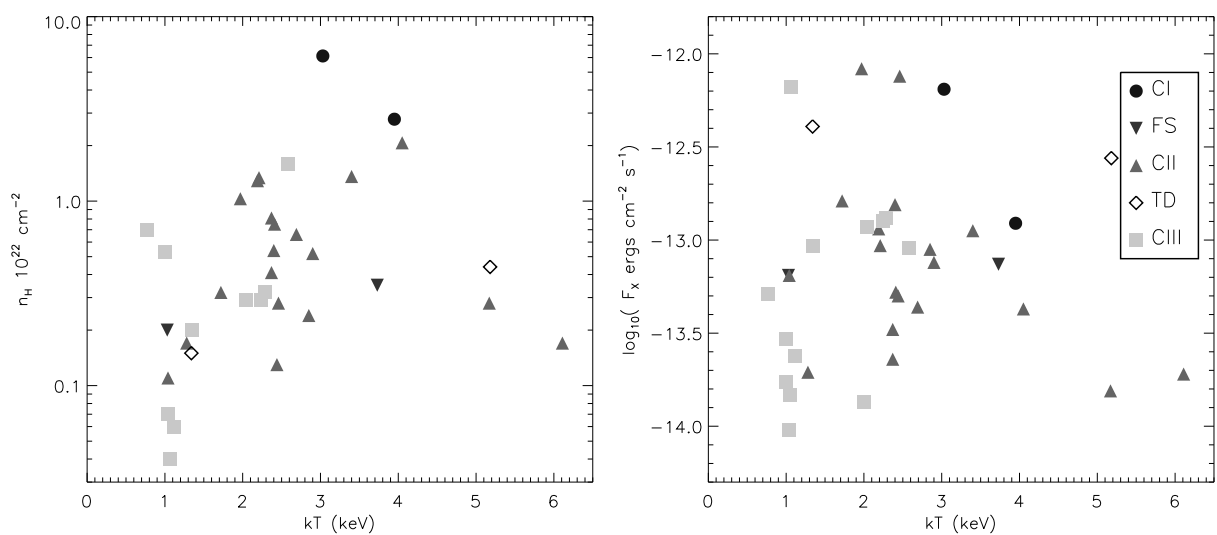}
\caption[NGC 1333 X-ray Characteristics]{ {\it Left:} Plasma temperature ($kT$) against Hydrogen 
 column density ($N_H$). The Class 0/I and 
 Flat Spectrum sources show higher $N_H$, consistent with the presence of an infalling envelope. 
 The Class II sources have higher median $N_H$ values than the Class III sources. 
 {\it Right:} The plasma temperature plotted against the log of the X-ray flux ($F_X$). A weak trend 
 of increasing flux with $kT$ is present. There is a significant difference in the value of $kT$ between 
 the Class IIs and Class IIIs.   The symbols are as follows: Class 0/I, circle; Flat Spectrum, inverted triangle; Class II, 
 triangle; Transition Disk, diamond; Class III, squares.   
}
\label{figxn}
\end{figure}

The observed X-ray detection limit can be estimated from \citet{fei} as  $log(L_{X}[erg s^{-1}]) \approx 27.6$, 
where the distance to NGC 1333 is taken as 240~pc, and the exposure time is 88.45~ks, and the assumed 
 $log(N_{H}[cm^{-2}])\approx22$.
Only those detections with an X-ray count in excess of 100 were considered when examining 
the following emission properties to ensure reliable estimates.
Fig.\ref{figxn} shows a plot of plasma temperature against the column density of hydrogen.   
There is a trend with evolutionary class in both $kT$ and $N_H$, with the Class I sources 
showing the highest values in both. The Class III sources exhibit the lowest values, 
from the lowest values we estimate the $N_H$ in the line of sight to the cluster from these diskless sources as 
$\sim0.1\times10^{22}$$cm^{-2}$.
The median values of hydrogen column density by class are: $N_H$ $= 6.12 \pm8.18 \times10^{22}$~$cm^{-2}$ 
for the Class I sources and Flat Spectrum sources,   $N_H$ $= 0.52 \pm0.52 \times10^{22}$~$cm^{-2}$ for the Class II 
and Transition Disk objects, and $N_H$ $=  0.20\pm0.43 \times10^{22}$~$cm^{-2}$ for the Class III stars.

Fig.\ref{figxn} shows the X-ray flux corrected for absorption and $kT$ for  the different 
classes.  The median flux of the three Class I sources falls in the same range as that of the Class II and III 
stars, suggesting that the protostellar and PMS X-ray flux levels are similar.   
\citet{jef} have shown increasing $kT$ with flux, and a trend toward cooler coronal temperatures in older clusters.   
Here, as in Serpens \citep{winston}, the evidence for an increasing trend in the $kT$ to $F_X$ is not clear, with a 
Spearman rank coefficient of 0.034.    If we discard the outliers, those with values  of 
$log(F_X[erg~cm^{-2}~s]) < -12.5$~$keV$ and $kT > 5.0$~$keV$,    
then we obtain a Spearman rank coefficient of 0.33, and a fit of $log(F_X)  \propto  0.47\pm0.07$~$kT$~$(keV)$, 
this is equivalent to  $log(F_X)  \propto  1.51\pm0.21 $~$kT$~$(MK)$.  The relation for the Serpens data is 
$log(F_X)  \propto  0.59\pm0.14$~$kT$~$(keV)$ ~$\propto  1.82\pm0.39 $~$kT$~$(MK)$.   
Previous studies have found values of $log(F_X) \propto  2.5$~$kT$~$(MK)$  for 1-Temperature model fits to late 
type stars \citep{schm} and $log(F_X) \propto  4.0$~$kT$~$(MK)$ for 2-Temperature model of ROSAT solar type 
field stars  \citep{gudel}.  \citet{prei97} finds a value of 2.2 for pre-main sequence stars in the Orion Nebula Cluster, 
similar to what we find.    We defer a more in-depth discussion of the plasma temperature and X-ray luminosities 
of the Class II and III sources to Section~\ref{ktsaef} where we perform a detailed analysis of the X-ray properties.

\subsection{Disk Fraction}

To estimate the number of missing diskless Class III objects in our sample, we make an initial 
assumption that the X-ray detection rates of Class II and Class III sources are similar, at 52\%.    
Further, from the examination of the spatial distribution and nearest neighbour comparisons, 
we find that, as in Serpens, the spatial distributions of the Class II and III sources are indistinguishable.  
Hence, as 41 Class III objects were observed, and taking into account that 94\% of the Class II population 
is contained in the {\it Chandra} field of view, the total Class III YSO population of the cluster may be 
estimated as $(41 / 0.52) / 0.94$, or 84 objects, which would bring the total cluster membership to 221.

The fraction of stars with disks in the region was calculated from the X-ray detected Class II and III 
stars, as $39/(41+39)$, or 49\%$\pm$8\%, similar to the 51\% found in Serpens. If the Class 0/I through 
Transition Disk objects are included then the fraction becomes: $(5+8+38+2)/(5+8+38+2+41)$, or 57\%$\pm$8\%.  
These fractions may be compared to the disk fraction calculated by \citet{gut07}, who subtracted the estimated number 
of background sources from the observed number of infrared sources in the NGC 1333 field with $K_S < 14$~mag. 
This led to an estimate of 87 YSOs in a circular region of radius 5.5$'$, covering the same central core as the
 {\it Chandra} field, of which 72 were IR-excess YSOs (Class I through Transition Disks), indicating a disk fraction 
 of $72/87$, or 83\%$\pm$11\%.  
Comparing our X-ray detections in the same circular region, we find 45 IR-excess YSOs and 27 Class IIIs, leading 
to a disk fraction of  63\%$\pm$9\%.  Adopting the same upper $K_S$-band magnitude limit at 14~mag as 
\citet{gut07}, we find 21 IR-excess YSOs and 15 Class IIIs, giving 58\%$\pm$13\%.    
The disk fraction estimated by \citet{wilk} of the northern filamentary cluster was 75\%$\pm$20\%. 
There is a hint that the X-ray selected samples give systematically lower disk fractions than those estimated 
by source counts, although the 1~$\sigma$ error bars of the results do (barely) overlap.
If the number counts method is yielding too high a disk fraction, this may be an   
underestimate of the number of Class IIIs due to an overestimate of the background contamination in the cluster.  
Conversely, the assumption that the fraction of Class III detected in X-rays is equal to the fraction of Class II detected 
may be erroneous.    This may be due to the somewhat higher X-ray luminosity in the Class III sources (see Sec.~\ref{fxlbol}.)

\section[XLFs]{X-ray Luminosity Functions: Distance to Serpens \label{xlfdist}}

\begin{figure}
\epsscale{.5}
\plotone{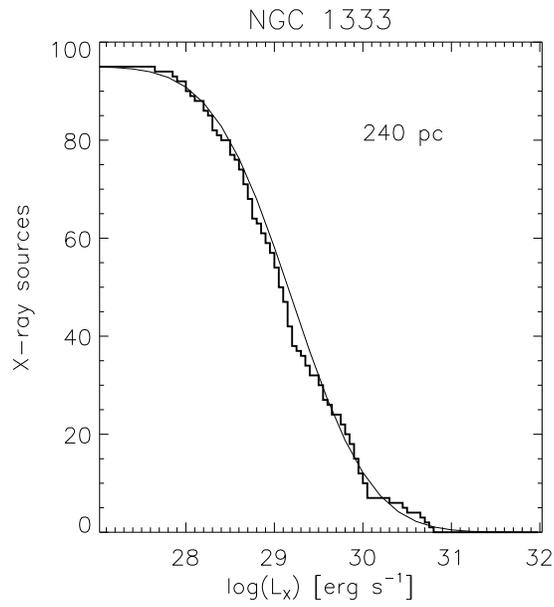}
\caption[X-ray Luminosity Function: NGC 1333]{ 
X-ray Luminosity Function of all {\it Chandra} detected sources for which flux measurements 
were determined in NGC 1333.   The data are fit to a log-normal distribution with 
$<log(L_X [erg s^{-1}])>  =  29.3$ and $\sigma = 1.0$, as determined by \citet{fei} for Orion, IC 348 and NGC 1333.  
 
}
\label{xlumn}
\end{figure}

The X-ray Luminosity Functions  (XLFs)  of young stellar clusters with ages $<$5~Myrs have been 
found to follow a universal distribution that can be fit by a lognormal with $<log(L_X[erg s^{-1}])>$ $=$  $29.3$  
and $\sigma_{log(L_X)} = 1$.  This has been demonstrated from consideration of observational data 
from the Orion COUP survey, IC 348, and NGC 1333 \citep{feiget,fei} and 
LkH$\alpha$101 \citep{wol2}.   The observed cluster XLF depends only on the number of sources  
and the distance to the cluster.  
In Fig.\ref{xlumn}  we plot the XLF of all the X-ray sources in NGC 1333 with $L_X$ calculated as 
in Sec.\ref{fxlbol}, using the maser parallax derived distance of 240~pc to the cluster \citep{hirota} and 
with the log-normal curve overplotted.  As was previously reported using  less sensitive {\it Chandra} 
ACIS-I imaging \citep{get2}, the NGC 1333 luminosity function is well fit by the universal log-normal, 
indicating that our adopted distance to NGC 1333 is likely to be accurate.

The distance to the Serpens Cloud is currently estimated as 260~pc \citep{strai96}.  
An early distance determination was that of \citet{strom} who measured the distance to HD 170734 
as 440~pc, assuming A0 spectral type, $V=$9.2 and $E_{B-V} \sim$0.3.  \citet{zha} reported 
distance of 700~pc was based on HD 170634, HD 170739, and HD 170784 with B spectral types, 
and $R=$3.1.    \citet{del} used the same stars, adding BD-24607 and Chavarria 7, reclassifying them 
to be on the main sequence, with $R_{BV}=3.3\pm0.3$ and obtained a distance of 310~pc.  
\citet{strai} details Vilnius photometry and photometric classification of 473 stars toward the Serpens 
Cauda cloud complex, to a depth of $V\sim13$.   The $A_V$ and distance to each star was calculated 
and used  to determine the near edge of the Serpens cloud and to estimate the depth of the cloud.  The 
near edge of the cloud is found to lie at 225$\pm$55~pc, with an estimated depth of 80~pc, leading to 
the average distance of 260~pc. The far edge of the cloud therefore lies at a maximum distance  
(assuming a +1$\sigma$ deviation) of $225+55+80=$ 360~pc.

\begin{figure}
\epsscale{1.}
\plotone{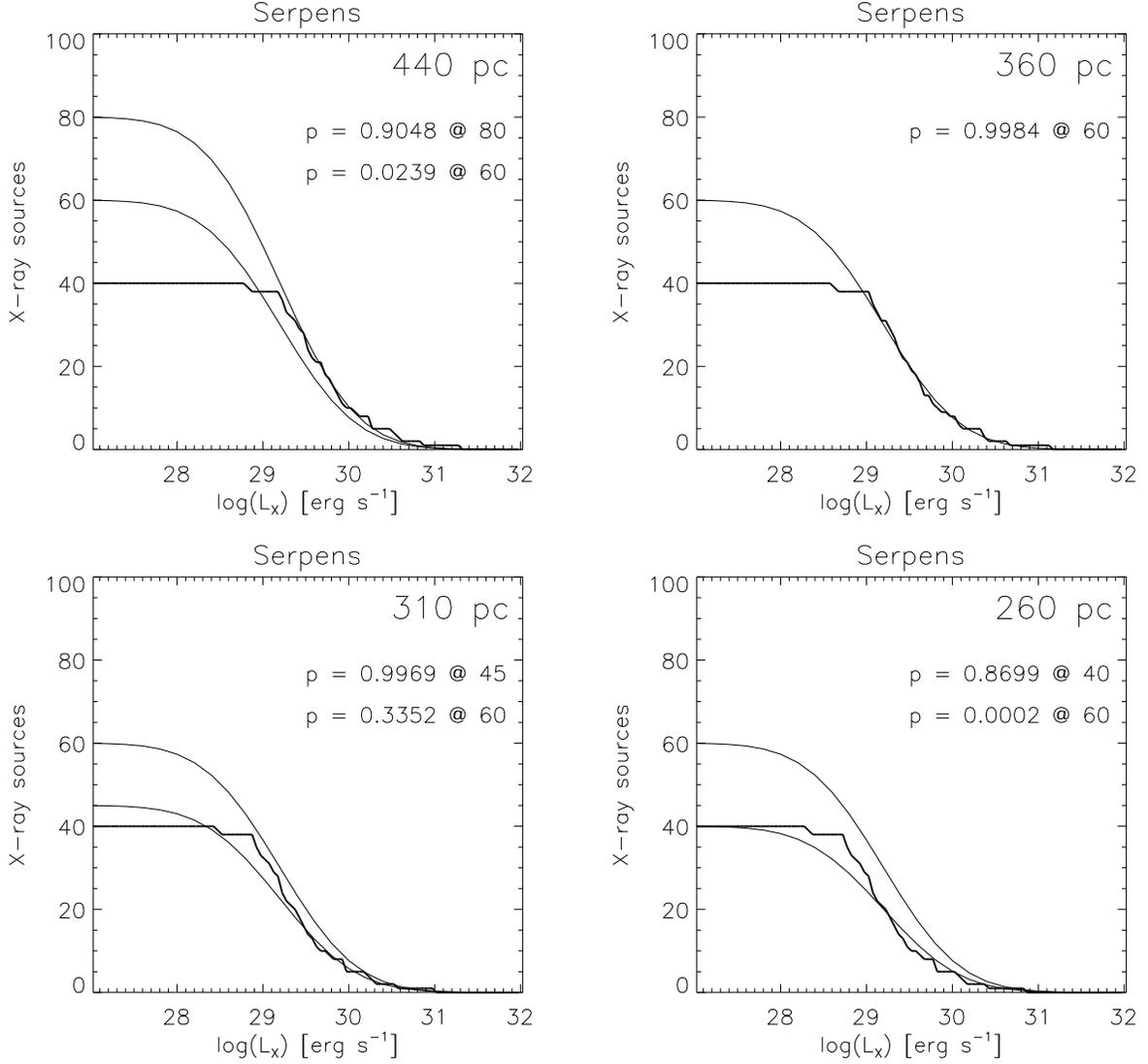}
\caption[X-ray Luminosity Function: Serpens]{ 
X-ray Luminosity Functions for the Serpens {\it Chandra} detected sources for which flux measurements 
were determined.  The XLFs are given for four different distances to the Serpens Core:  440pc, 310pc, 
260pc (previous distance estimates from the literature, see text), and 360pc, the best fit to the log-normal 
distribution from these data.  The data were fit to a log-normal distribution with $<log(L_X [erg s^{-1}])>$ $=   29.3$ 
and $\sigma = 1.0$, as determined by \citet{fei} for Orion, IC 348 and NGC 1333.   The fit is performed on 
the tail of the distributions above $log(L_X [erg s^1])$ $= 29.3$ where the luminosity bins are considered 
to be complete and the universal XLF is best defined.    The minimum $\chi^2$ fit is overplotted  for the case 
where the number of X-ray detected objects is a free parameter and for the case where the number of 
objects is constrained to 60 sources.   The probabilities associated with both of these fits is also given.  
 
}
\label{xlums}
\end{figure}

In Fig.\ref{xlums}  we plot the XLFs for the Serpens X-ray sources from \citet{winston} with modelled fluxes 
using four different distances to calculate the luminosity.  Three of these are from previous literature 
studies: 260~pc (the currently most commonly adopted distance), 310~pc, and 440~pc, 
and one at 360~pc.  We performed a least squares fit of the observed distribution to the universal XLF 
of \citet{get}, varying the number of sources between 39 and 81 members and incrementing the distance 
by 10~pc from 250 to 450~pc.  The fit was performed on the tail of the distribution from 
$log(L_X)$$\ge$$29.3$ to avoid including any incomplete luminosity bins in the sample and to fit to 
the universal XLF where it is best defined.   
The figure shows the minimum $\chi^2$ fit at each distance and the fit for 60 sources, the $\chi^2$ probabilities 
for both fits are also given.   
The best fit to the data was found to be at a distance to Serpens of $360^{+22}_{-13}$~pc and 60 detections, 
with a $\chi^2$ probability of 0.998.   
To estimate the uncertainties, we determine the distances at which  $\chi^{2} = min(\chi^{2}) \pm 1$ while 
constraining the number of objects to 60 and varying the distance in steps of 1~pc.   
There are 60 {\it Chandra} detections coincident with IR sources in the Serpens field, though only 40 had 
high enough count rates to have luminosities determined through model spectrum fits (hence the flat topped 
observed luminosity distributions in Fig.\ref{xlums}).   As such, the model fit to sixty sources appears to be reasonable.  
We note that if the number of X-ray luminous members is in excess of 60, higher distances are likely.  For example, 
a high probability of 0.9 is found for 40~pc {\it if} we allow 80 objects - 20 more than have been detected.  In contrast, 
distances of 310~pc or lower require the numbers of X-ray luminous members to be {\it less} than the number detected 
and may be ruled out.   
This best fit distance is approximately mid-way between the \citet{del} 310~pc and \citet{strom} 440~pc 
estimates, and indicates that Serpens may be further than the commonly adopted 260~pc.  
While statistical uncertainties and systematic errors in the determination of $L_X$ will introduce scatter and 
cause uncertainty in the XLF fitting, and the universal XLF itself is not exactly lognormal, these errors 
should apply equally to NGC 1333. 
As NGC 1333 is consistent with the universal XLF for the  known distance and cluster membership, 
we find it unlikely that the more distant fit of Serpens can be due to these effects alone.   
Given that similar velocities have been found for the molecular gas  associated with the recently discovered 
Serpens South cluster \citep{gut08} and that associated with the Serpens Cloud Core studied here, we 
suggest that this further distance apply to the southern region as well.  

\begin{figure}
\epsscale{0.8}
\plotone{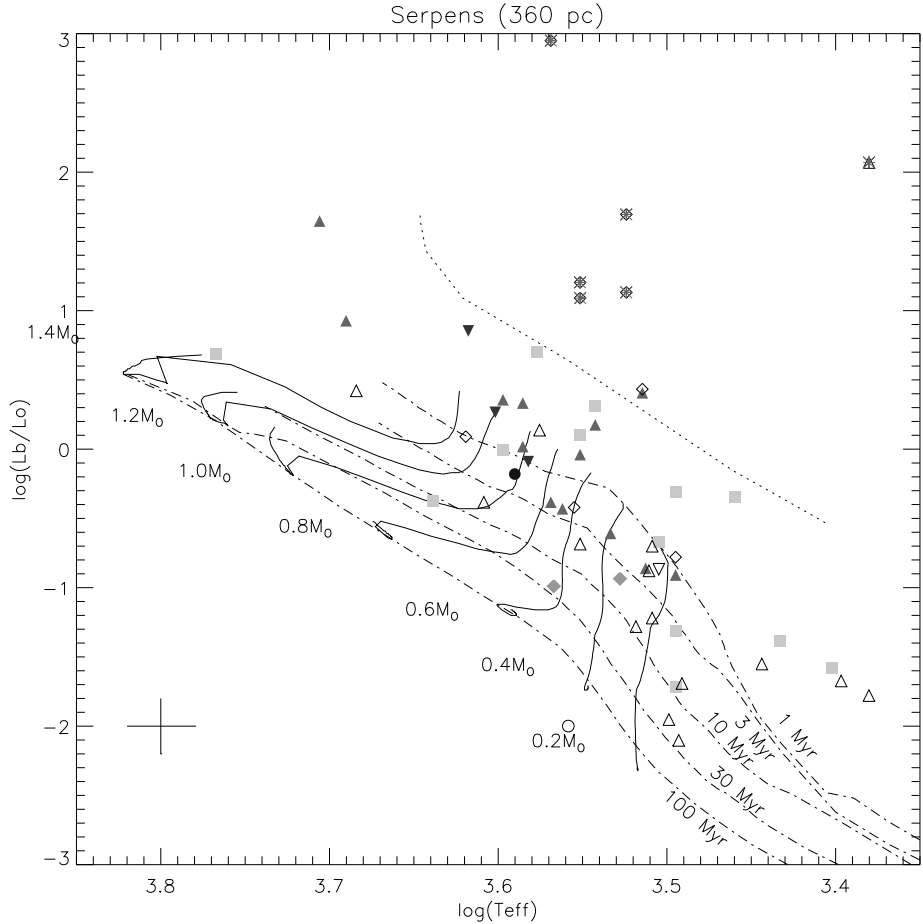}
\caption[HR diagram: Serpens by class \& X-ray detection]{ 
The H-R diagram of the Serpens cluster plotting all stars with well established spectral types, 
using a distance of 360~pc to calculate the luminosity.  
The sources are identified by their evolutionary classification, with circles indicating 
the Class 0/I protostars, the inverted triangles the Flat Spectrum.   The Class II objects are 
represented by triangles, Transition Disks by open diamonds, and Class III sources 
by squares.   The filled symbols plot those sources with {\it Chandra} X-ray counterparts, 
the empty symbols, those without.  The evolutionary tracks are taken from \citet{bar}, with the 
1, 3, 10, 30, and 100~Myr isochrones, and mass tracks from 0.2 to 1.4~$M_{\sun}$.  
The stellar birthline is plotted as a dotted line \citep{dan}.  The six highly luminous objects, 
four Transition Disks and one Class II object without X-rays, are very likely contaminating AGB 
stars, and are marked with asterisks \citep[see][for a discussion of these objects]{win08a}.   
The error bars for typical uncertainties are shown in the lower left corner.    
}
\label{hrdscx}
\end{figure}

To ascertain whether this new estimated distance is compatible with our previous study of the Serpens HR 
diagram (HRD), in Fig.\ref{hrdscx} we have replotted the HRD of the region using the distance of 360~pc to 
calculate the luminosities of the spectrally classified YSOs  (c.f. Fig.6 in \citet{win08a} for the HRD at 260~pc).    
At the new distance, the median age of both the Class II and III YSOs decrease to 1~Myrs.  Two objects are 
newly located at or above the birth line, and these may be  over-luminous, perhaps due to binarity,  
variability, or uncertainties in the position of the birthline.  The majority of sources now have isochronal ages less than 3~Myr.  
At this distance, only four members now have apparent isochronal ages $>$10~Myr and none have ages $>$30~Myr.   
These ages are more in keeping with what might be expected of YSOs in a deeply embedded cluster rich with 
protostars.    In the following discussions in this paper, both the bolometric and the X-ray luminosities for the 
Serpens cluster have been calculated from this new distance of 360~pc.       

\begin{figure}
\epsscale{1.}
\plotone{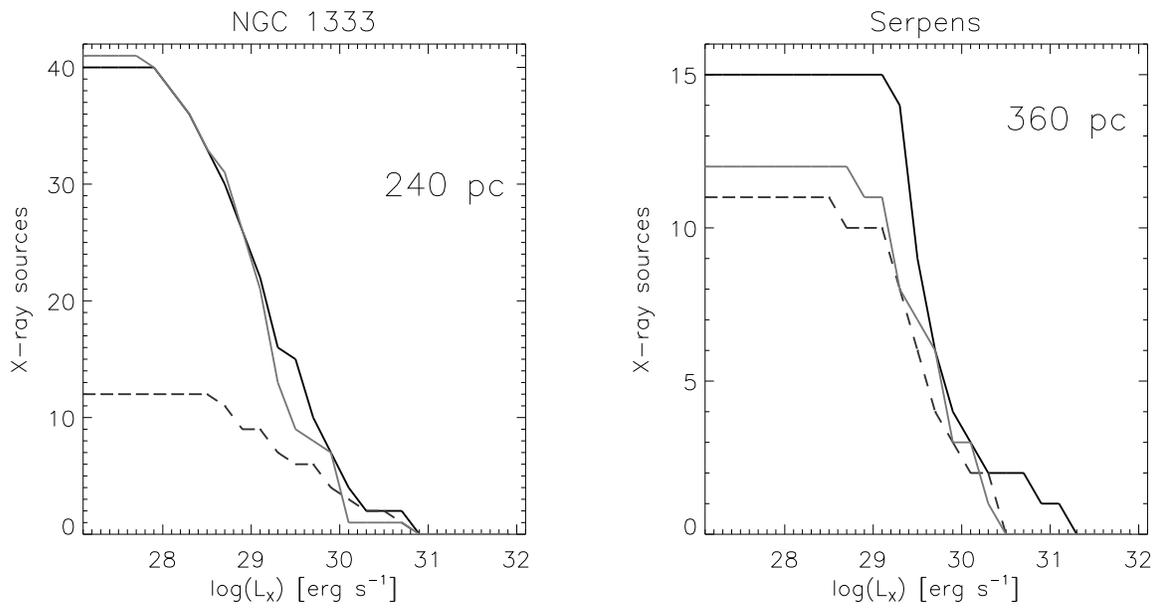}
\caption[X-ray Luminosity Function by Evolutionary Class]{ 
X-ray Luminosity Functions of NGC 1333 and Serpens for sources with absorption corrected luminosities.  
The Class II sources are shown by the solid black line, the Class III by the grey line.  
The dashed black line indicates the protostars (Class 0/I and Flat Spectrum). 
In both clusters, the K-S probabilities suggest that the three groups are from the same parent distribution.

}
\label{xlumcl}
\end{figure}

Does the XLF depend on evolutionary class? Fig.\ref{xlumcl} shows the XLFs of the two clusters by evolutionary 
class for the Class II, Class III, and protostellar sources (Class I and Flat Spectrum).  In both clusters the Class II 
and Class III members show similar XLFs, with a Kolmogorov-Smirnov test probability of their arising from the same 
parent distribution of 48\% in NGC 1333 and 65\% in Serpens, for sources with $log(L_X [erg s^{-1}]) \ge 29$ in Serpens 
and $\ge 28.5$ in NGC 1333.  These cut-offs in luminosity are imposed so that the K-S tests are based on complete bins 
in all evolutionary groups.  The K-S probabilities indicate that the detection rate of the Class III sources is comparable to 
that of the Class II.   This is similar to \citet{get2}, who found no differences in the XLFs of the Class II and Class III sources 
in the ONC.   In contrast,  \citet{tell} show WTTS to be brighter in X-rays than CTTS in Taurus.  In Section~\ref{fxsate} we 
will further discuss the differences in Class II and III X-ray luminosities.   

The protostellar sources also show similar XLFs to those of the more evolved objects, although the sample is biased 
towards higher luminosities by their higher extinctions.  In NGC 1333, the XLFs have probabilities of 71\% and 34\%  
of being drawn from the same parent distribution as the Class II and III sources, respectively.  In Serpens, those 
probabilities are 99\% and 86\% for Class II and III sources, respectively.  Thus, we find no evidence that the protostellar 
luminosity function differs from the primarily Class I and Flat Spectrum sources.

\section[]{The Dependence of X-ray Emission on Measured Stellar Properties}

\subsection[$L_X$ v. $L_{bol}$]{X-ray Luminosity and Bolometric Luminosity \label{fxlbol}}

While the X-ray luminosity of main sequence stars, like the Sun, is variable, it also scales as a function 
of the bolometric luminosity and is dependent on the Rossby number (a measure of the stellar rotation rate) 
and the convective turnover time \citep{noy}. The X-ray luminosity of pre-main sequence stars has also been 
found to vary with bolometric luminosity \citep{fei4,pre}.  Coronal magnetic activity appears to be less variable 
in pre-MS stars that in MS stars \citep{ste}, and is likely the origin of most of the X-ray emission from the YSOs, 
though accretion processes may also have an effect.   \citet{pre} and \citet{pre2002} examined the relation of 
$L_X$ to $L_{bol}$ for YSOs in the Orion COUP survey and IC 348, and for a sub-sample of field stars from the 
NEXXUS survey.  They find that for pre-main sequence stars $L_X \propto L_{bol}^{\alpha}$, where $\alpha$$\sim$1, 
compared to  $\alpha$$\sim$0.4 for stars on main sequence.  This relation was also observed for pre-main 
sequence stars in the Taurus molecular cloud by the XEST survey \citep{tell}.

We applied a similar analysis, determining the X-ray luminosity from $L_X  =   4 \pi D^{2}  F_X$  where $D$ 
is the distance (we use $F_X$ tabulated in Table 1 for NGC 1333 and on the ANCHORS website for Serpens, using 
our new estimate of the distance in the calculation), and using the $L_{bol}$  from \citet{winston}.   
Tables \ref{tableserprop} \& \ref{tablen13prop} list the spectral, X-ray and IR properties of the 138 YSOs 
considered in the following studies.  The tables list the {\it Spitzer} identifier, spectral type, evolutionary 
class, {\it Chandra} identifier, bolometric luminosity, effective temperature, stellar surface area, the X-ray 
luminosity, plasma temperature and hydrogen column density, the extinction at $K$-band, isochronal age and 
mass.   The Serpens data are taken from \citet{winston} and \citet{win08a}, while the NGC 1333 IR and spectroscopic 
data are taken from \citet{gut08} and \citet{win08a}, and the NGC 1333 X-ray data from this work.  
The isochronal ages and masses were determined from the isochrones and mass tracks of \citet{bar} 
interpolated onto a grid using IDL routines; the earliest isochrone used was 1~Myr.  Those sources with isochronal ages 
listed as $\sim$0.2~Myr were near  to the boundary of the grid used for extrapolation, where the age estimates were less 
certain, and were therefore assigned an estimated isochronal age.  Those without an isochronal age or mass measurement 
were located beyond the boundaries of the grids where a useful age/mass estimate could not be obtained.   
The extinction at $K$-band was measured following \citet{gut08} - who uses the CTTS locus first discussed by \citet{mey}.  
Where a value of $K$-band extinction could not be determined or was found to be blueward of the CTTS locus, the 
$A_{K}$ was set to zero in the calculations.   In Tables \ref{tableserprop} \& \ref{tablen13prop} an $A_K$ of 0. indicates 
that the source was blueward of the CTTS locus, and a null value that it could not be calculated.

\begin{figure}
\epsscale{1.}
\plotone{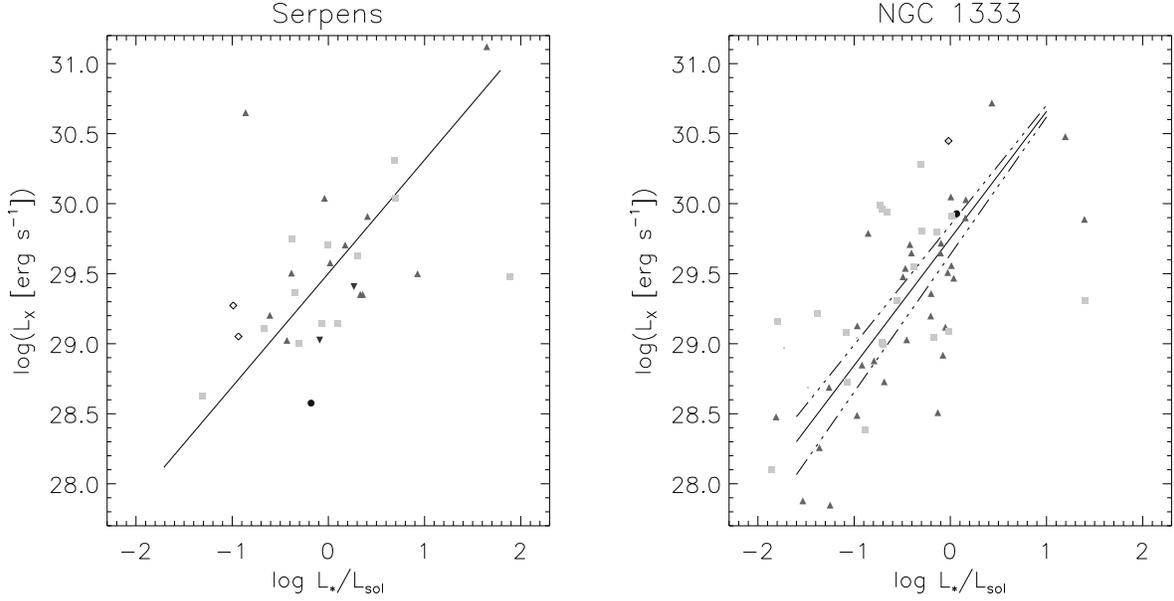}
\caption[Stellar Bolometric to X-ray Luminosity]{
 In the graphs, the X-ray luminosity, $L_X$, is plotted against the bolometric luminosity, 
$L_{\star}$/$L_{\sun}$, for the Serpens (left) and NGC 1333 (right) clusters. A linearly 
increasing trend is observed in both,  with slopes consistent with previous observations 
of the Orion Nebular Cluster and IC 348.  
Symbols indicate Class 0/I (circle), Flat Spectrum (inverted triangle), Class II (triangle), 
Transition Disk (diamond), Class III (square).  
The dot-dashed lines indicate the best fit to the entire YSO sample in each region, with 
 $log(L_X [erg s^{-1}])$ $\propto$~$({\bf 0.82 \pm 0.11})$~$log(L_{*}/L_{\odot})$ in Serpens and 
 $log(L_X [erg s^{-1}])$ $\propto$~$({\bf 0.91 \pm 0.10})$~$log(L_{*}/L_{\odot})$ in NGC 1333. 
The dashed lines on the NGC 1333 plot are the fits to the Class III sources (upper line: 
$log(L_X [erg s^{-1}])$ $= 29.85\pm0.19$~$+$~$(0.86 \pm 0.14)$~$log(L_{*}/L_{\odot})$) and 
Class II sources (lower line: $log(L_X [erg s^{-1}])$ $= 29.69\pm0.10$~$+$~$(0.98 \pm 0.13)$~$log(L_{*}/L_{\odot})$). 

}
\label{xlumbol}
\end{figure}

Figure~\ref{xlumbol} plots the stellar X-ray luminosity against the stellar bolometric luminosity in units of 
$L_{\sun}$ for the Serpens (left) and NGC 1333 (right) clusters.    In both regions, a trend of increasing 
$L_X$ with stellar bolometric luminosity is observed.
The slope of the relationship in the Serpens data is:  
 $log(L_X [erg s^{-1}])$ $\propto$~$({\bf 0.82 \pm 0.11})$~$log(L_{*}/L_{\odot})$, 
and the slope  in the NGC 1333 region is:
 $log(L_X [erg s^{-1}])$ $\propto$~$({\bf 0.91 \pm 0.10})$~$log(L_{*}/L_{\odot})$.   
In both regions, the bolometric values correspond, within 2 $\sigma$, to the 
$L_X \propto L_{bol}$ relation observed previously in other young stellar clusters.  
For the NGC 1333 region, the Class II sources appear slightly less luminous than the Class III, but with 
the large scatter in the data the difference is less than 1~$\sigma$.  
The power law exponent for the Class II sources is:
$log(L_X [erg s^{-1}])$ $= 29.69\pm0.10$~$+$~$(0.98 \pm 0.13)$~$log(L_{*}/L_{\odot})$,
while for the Class III sources it is:
$log(L_X [erg s^{-1}])$ $= 29.85\pm0.19$~$+$~$(0.86 \pm 0.14)$~$log(L_{*}/L_{\odot})$.

In \citet{winston}, a discussion of the stellar X-ray flux to the dereddened $J$-band magnitude was put forth.  
The near-IR magnitude was used as a proxy for the stellar bolometric luminosity.   
In Serpens and NGC 1333, respectively, we find values of $F_X \propto m_{J}^{\alpha}(dered)$ with $\alpha$ 
of 0.42$\pm$0.17 and 0.53$\pm$0.09, where $m_{J}(dered)$ is the dereddened magnitude.   
\citet{cas}, working from the results of \citet{gre}, found empirically that $log(L_{bol}/L_{\sun}) \propto -0.4m_{J}(dered)$.  
Therefore, the $F_X$ vs. $m_J(dered)$ relationships are consistent with our derived $L_{bol}$ and $L_X$ relationships.

\subsection{$L_X$ with Surface Area \& Effective Temperature}\label{fxsate}

\begin{figure}
\epsscale{1.}
\plotone{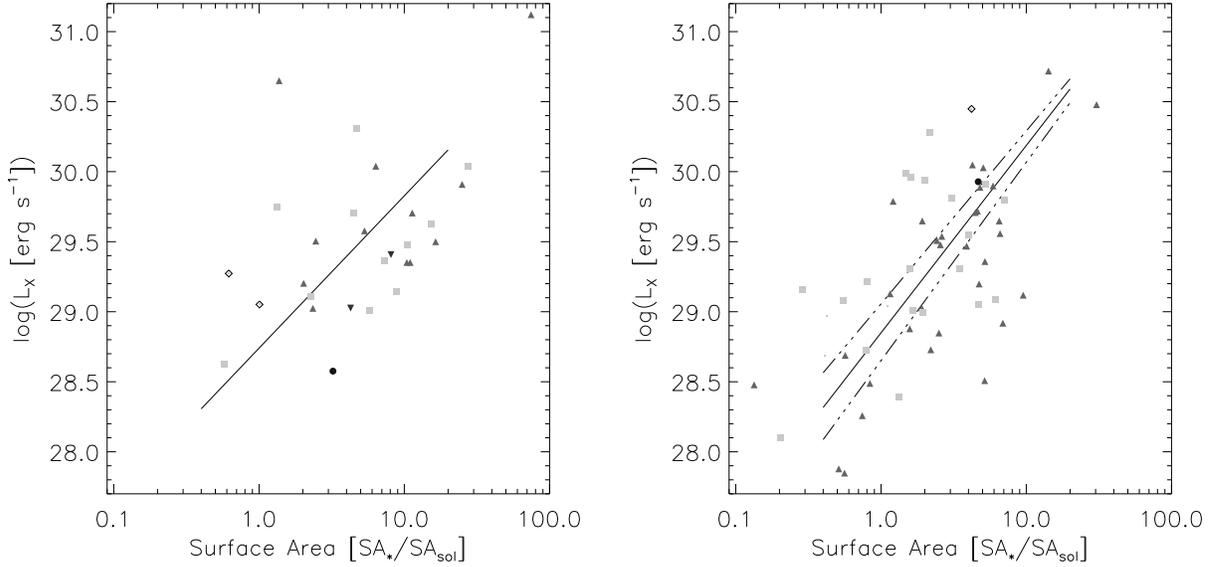}
\caption[Surface Area to X-ray Luminosity]{
In the two plots, the relation of X-ray luminosity, $L_X$, to the stellar surface area is plotted for 
Serpens (left) and NGC 1333 (right).  There is a trend of increasing surface area with luminosity, 
where $L_{bol}$ $\propto$ $4\pi R^{2}$.  
Symbols indicate Class 0/I (circle), Flat Spectrum (inverted triangle), Class II (triangle), 
Transition Disk (diamond), Class III (square).   
The dot-dashed lines indicate the best fit to the entire YSO sample in each region, with
$log(L_X [erg s^{-1}])$ $= 28.74\pm0.13$~$+$~$({\bf 1.09 \pm 0.10})$~$log(SA_{*}/SA_{sol})$ in Serpens and 
 $log(L_X [erg s^{-1}])$ $= 28.85\pm0.08$~$+$~$({\bf 1.34 \pm 0.12})$~$log(SA_{*}/SA_{sol})$ in NGC 1333.
The dashed lines on the NGC 1333 plot are the fits to the Class III (upper line) and 
Class II (lower line) subsamples, showing the Class IIIs marginally brighter in $L_{X}$ than the Class II for a given 
$S.A.$, with
$log(L_X [erg s^{-1}])$ $= 29.06\pm0.10$~$+$~$(1.24 \pm 0.16)$~$log(SA_{*}/SA_{sol})$, for the Class III and 
$log(L_X [erg s^{-1}])$ $= 28.65\pm0.13$~$+$~$(1.42 \pm 0.18)$~$log(SA_{*}/SA_{sol})$, for the Class II. 
}
\label{xlumsurfarea}
\end{figure}

\begin{figure}
\epsscale{1.}
\plotone{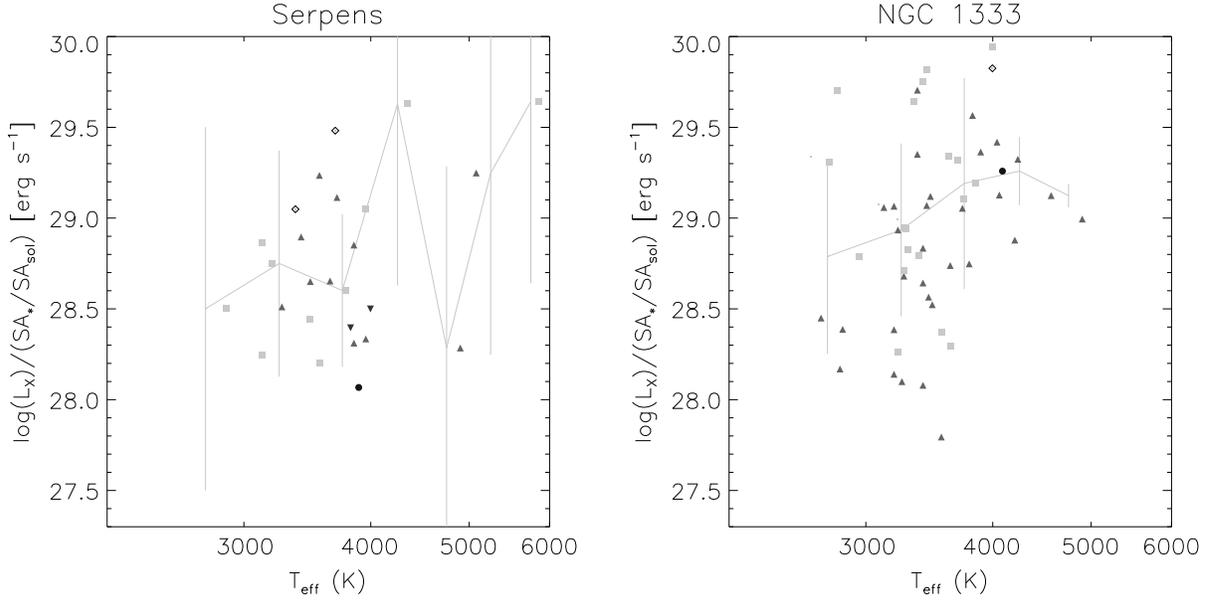}
\caption[Effective Temperature to X-ray Luminosity/S.A.]{
 In the two graphs, the X-ray luminosity, $L_{X}/(SA_{*}/SA_{sol})$, is plotted against the stellar effective 
 temperature, $T_{eff}$, for the Serpens (left) and NGC 1333 (right) clusters. To remove the 
 effect of surface area, this has been divided out of $L_X$.   
A weak increase of $L_{X}/(SA_{*}/SA_{sol})$ with $T_{eff}$ is apparent in NGC 1333, with 
a jump in $L_{X}/(SA_{*}/SA_{sol})$ between M0 and K7 spectral types at 3800~K.   
Symbols indicate Class 0/I (circle), Flat Spectrum (inverted triangle), Class II (triangle), 
Transition Disk (diamond), Class III (square).   
The overplotted dashed lines show the median $L_{X}/(SA_{*}/SA_{sol})$ in bins of 500~K 
in $T_{eff}$ from 2500-6000~K. The vertical lines give the standard deviation.   
}
\label{xlumteff}
\end{figure}

In this section we will examine the dependence of $L_X$  with surface area and with $T_{eff}$ to 
discover if the increasing trend of X-ray flux to bolometric luminosity arises from variations in the 
effective temperature or the surface area of the star.  
The bolometric luminosity of a star may be given as $L_{bol} =  4\pi R^{2} \sigma T_{eff}^{4}$, where 
$4\pi R^2$ is the stellar surface area and $T_{eff}$ is the effective temperature.     
The surface area was determined in solar units from the bolometric luminosity and effective 
temperature derived from the spectral classification of each star, determined in \citet{win08a}.  
The plots in Figs.~\ref{xlumsurfarea} \& \ref{xlumteff} examine the physical parameters on 
which the $L_X$ vs. $L_{bol}$ relation may depend: stellar surface area, $S.A.$, and effective temperature, $T_{eff}$.    
We note that the stellar surface area and effective temperature are not independent parameters: they will show some 
interdependence, such that hotter, more massive pre-main sequence stars will on average have larger surface areas.

In Fig.~\ref{xlumsurfarea}, the X-ray luminosity is plotted against the stellar surface area: 
The trend to increasing $L_X$ with stellar surface area is clearly visible in the plots.  
In Serpens, the fit to these data is 
 $log(L_X [erg s^{-1}])$ $= 28.74\pm0.13$~$+$~$({\bf 1.09 \pm 0.10})$~$log(SA_{*}/SA_{sol})$, 
while in NGC 1333 it is   
 $log(L_X [erg s^{-1}])$ $= 28.85\pm0.09$~$+$~$({\bf 1.34 \pm 0.12})$~$log(SA_{*}/SA_{sol})$.  
In Serpens, the Class II and III YSOs do not show any difference in trend of $L_X$ to $log(SA_{*}/SA_{sol})$.   
In NGC 1333, for a given surface area, the Class III YSOs have a somewhat higher X-ray flux than the Class IIs.  
The two populations can be fitted separately, with 
$log(L_X [erg s^{-1}])$ $= 28.65\pm0.13$~$+$~$(1.42 \pm 0.18)$~$log(SA_{*}/SA_{sol})$ for the Class III and 
$log(L_X [erg s^{-1}])$ $= 29.06\pm0.10$~$+$~$(1.24 \pm 0.16)$~$log(SA_{*}/SA_{sol})$ for the Class II.  
This difference in the median luminosity of the two fits is approximately 0.3~dex.  
Thus, we find a marginal offset of about $2\sigma$ between the fits of the two evolutionary classes.
It is not clear why Serpens and NGC 1333 show different results.  We do note that the Serpens 
sample is smaller (12 Class II and 11 Class III vs.  31 Class II and 23 Class III in NGC 1333) and dominated 
by more X-ray luminous objects where the difference between Class II and III objects is less obvious; hence
such a difference could be less apparent.   

In Fig.~\ref{xlumteff}, the X-ray luminosity is compared with the effective temperature, $T_{eff}$. 
In order to remove the contribution from surface area and thus isochronal age, the surface area 
has been divided out of the X-ray luminosity data to give the surface flux: $L_{X}/(SA_{*}/SA_{sol})$.   
The overlaid graphs plot the median value of $L_{X}/(SA_{*}/SA_{sol})$ in 
bins of 500~K in $T_{eff}$ from 2500~K to 6000~K.  There is a discontinuity in the NGC 1333 plot, 
 which occurs at an approximate temperature of 3800$K$, the transition 
between $M0$ and $K7$ spectral types.   A Kolmogorov-Smirnov test was performed on the 
populations to either side of this boundary temperature to ascertain the probability that they are from 
the same parent distribution: in NGC 1333 the probability was 7.4\%, in Serpens 29.3\%.   
Although the result in NGC 1333 is suggestive, a larger sample of sources is needed to provide the 
statistical significance necessary to confirm the discontinuity.

Previous studies have arrived at contradictory answers on the issue of the X-ray luminosity of Class II 
and Class III YSOs.  
A number of studies have found no difference in the X-ray luminosities of the two groups 
\citep{gag,fei4,cas,pre2002}, while in Taurus, \citet{stel} and \citet{tell} find the WTTS to be more luminous. 
\cite{pre} find that while accreting T Tauri stars are less luminous than non-accretors, 
those stars with $K$- or $L$-band excesses show no difference in luminosity to those without.   
It is not clear how much of this is due to the intrinsic scatter in the X-ray luminosities  
(which is greater than the systematic differences between Class II and III sources), real differences between 
different star forming regions, and the different criteria (accretion signatures and near IR excess emission) 
used to separate Class II and Class III.   \citet{win08a} found a strong correspondence between the presence 
of an accretion indicator and the presence of an IR-excess from a disk; thus, we do not expect a difference in 
behavior between samples selected by accretion indicators and those selected by IR-excess.  The fact that 
even with the {\it Spitzer} data, we find one cluster with differing Class II and II X-ray luminosities (NGC 1333) 
and one without (Serpens) suggests that the contradictory results may result from a combination of the large 
intrinsic scatter and the smaller sample size in Serpens.

\begin{figure}
\epsscale{1.}
\plotone{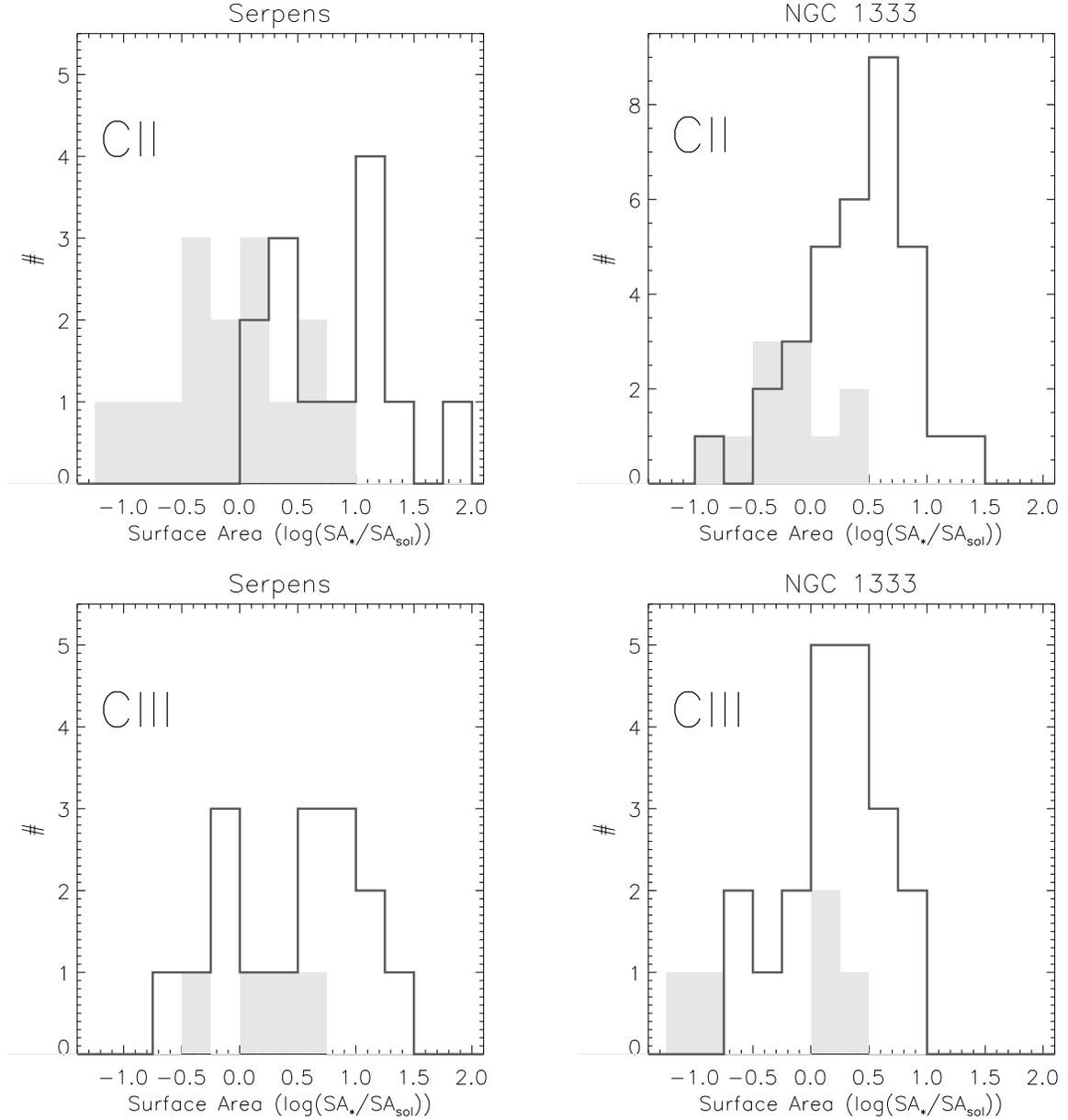}
\caption[Surface Area Histograms of Xray/Non-Xray YSOs.]{
 Histograms of the stellar surface area, $SA_{*}/SA_{sol}$, for the Serpens (left) and NGC 1333 (right) 
 clusters.  The upper plots show the Class II sources, the lower plots the Class III.  The 
 solid black lines indicate the X-ray detected YSOs.  The grey shaded histograms are those 
 of the non X-ray detected YSOs.   For the Class IIIs, the non X-ray detections are those identified 
 by $Li~I$ absorption in their spectra, c.f. \citet{win08a}.   Those sources detected in X-rays have, 
 on average, larger surface areas than those not detected.   
 }
\label{surfXnX}
\end{figure}

Given the dependence of $L_X$ on surface area for detected sources, it is also important to assess the 
detection rate of YSOs as a function of their surface areas.     
In Fig.\ref{surfXnX}, the histograms of the surface areas of the Class II and Class III YSOs with 
spectral types in each cluster are presented for both clusters.  The solid black histograms indicate 
the X-ray detected YSOs, while the shaded grey histograms are those of the non X-ray detections.   
The Class II sources without X-ray detections are easily detected by their IR-excesses.  The 
displayed sample are the Class II sources for which we obtained spectra; \citet{win08a} describe the 
selection biases in that sample.   
The histograms show that the X-ray detected Class IIs have larger surface areas than the 
non X-ray detections, with median values of 2.62 and 0.61~$S.A._{\odot}$ for NGC 1333 and 3.32 and 
0.70~$S.A._{\odot}$ for Serpens, respectively.   The Class IIIs have median values of 1.94~$S.A._{\odot}$ 
for NGC 1333 and 2.35~$SA_{*}/SA_{sol}$ for Serpens.   To attempt to find a value for the non X-ray detected 
Class IIIs we have utilised the candidate Class IIIs identified from $Li~I$ in their spectra \citep{win08a}.   
Their median surface areas were 1.11 and 1.54~$SA_{*}/SA_{sol}$ for NGC 1333 and Serpens, respectively, though 
the statistics here are small.  Not only are the median surface areas of the non-detected sources smaller,  
but we find that all of the non-detected sources have  $log(SA_{*}/SA_{sol}) < 1$ for Serpens and $<0.5$ for NGC 1333.      
Such sources with smaller surface areas may emit too weakly in X-rays to be detected in our {\it Chandra} observations 
unless they are undergoing strong flaring.

\subsection{Plasma Temperature with Surface Area and Effective Temperature}\label{ktsaef}

\begin{figure}
\epsscale{1.}
\plotone{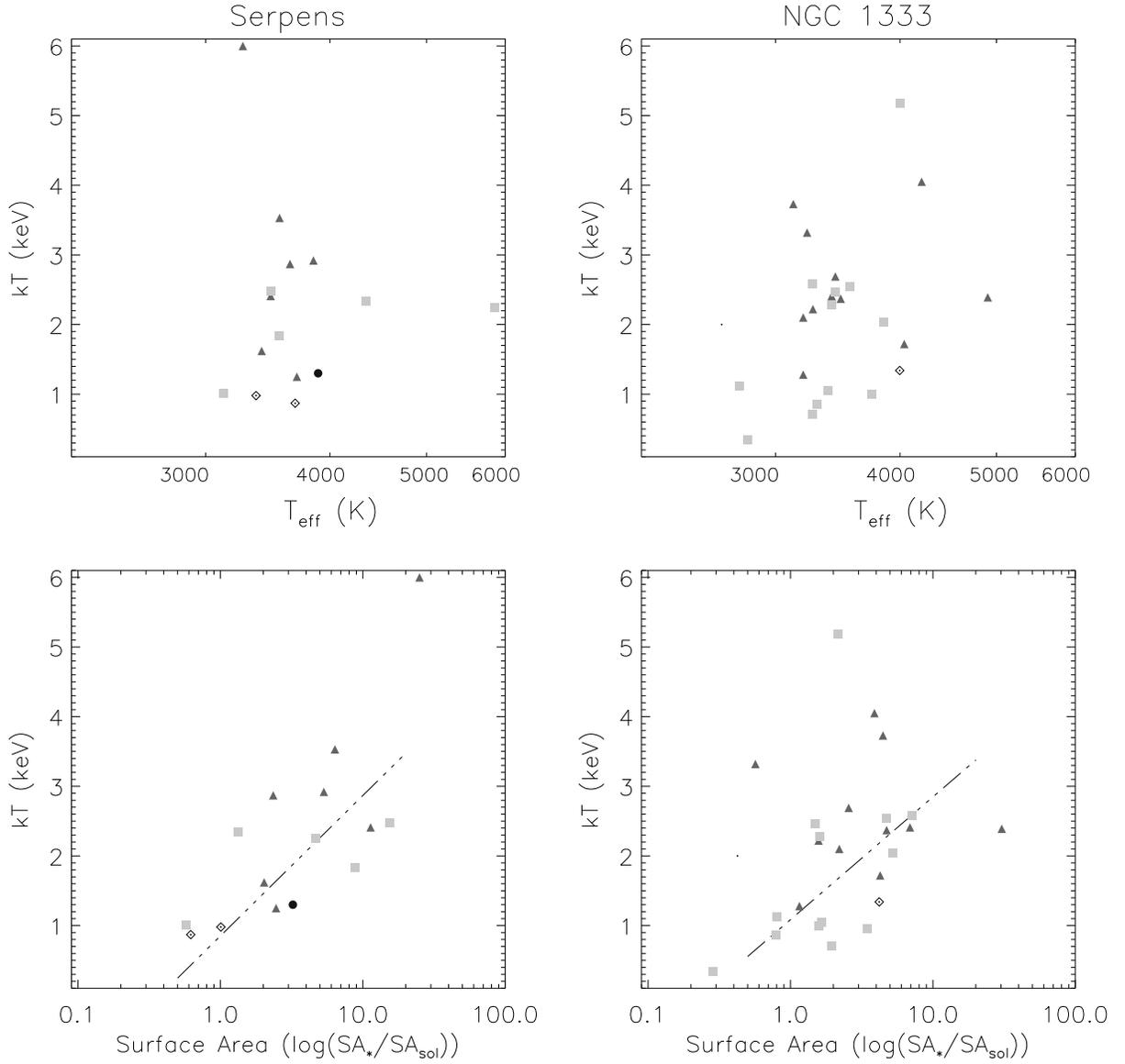}
\caption[Stellar Surface Area \& Effective Temperature to Plasma Temperature]{ 
The upper plots show $T_{eff}$ against the X-ray plasma temperature, $kT$.  
The lower plots show the surface area against $kT$.  A trend of increasing plasma 
temperature with  surface area is noted. 
Symbols indicate Class 0/I (circle), Flat Spectrum (inverted triangle), 
Class II (triangle), Transition Disk (diamond), Class III (square).   
The dash-dot lines show the fits to the data, indicating an increasing trend in $kT$ with 
surface area, of   $kT [keV]$$= 0.85\pm0.22$~$+$~$(2.02\pm 0.28)$~$log(SA_{*}/SA_{sol})$ in Serpens 
and  $kT [keV]$$= 1.09\pm0.26$~$+$~$(1.76\pm 0.22)$~$log(SA_{*}/SA_{sol})$ in NGC 1333.  
The fits to these plots were made while removing the outlier data points:  in Serpens where 
$kT$$\approx$6~keV, and in NGC 1333 where  $kT$$>$3~keV.   

}
\label{surftemp}
\end{figure}

In Section~\ref{xrayc}, the X-ray luminosity was found to increase with the plasma temperature. 
A hotter plasma may arise from increased heating rate in the coronal loop or through increased 
loop length \citep{prei97}.   \citet{gag} note that the plasma temperature is dependent on the 
stellar rotation period and stellar age.   \citet{pre} also find that the plasma temperature increases 
with $T_{eff}$. 
As $L_{X}$ is dependent on $SA$ and $T_{eff}$,  in Fig.~\ref{surftemp}, we   
investigate the dependence of the X-ray plasma temperature, $kT$, on the surface area and 
effective temperature of the star.  The fits to these plots were made while removing the outlier 
data points:  the outlier points are at $kT$$\approx$6~keV in Serpens, and at $kT$$>$3~keV 
in NGC 1333 (see the $kT$ vs $SA$ plot in Fig.~\ref{surftemp} for the motivation for removing 
these points.)    
Neither cluster shows strong evidence of a trend in plasma temperature with increasing effective temperature. 
Both clusters do show  trends of increasing plasma temperature with increasing surface area.   
The Serpens data show a trend of  
 $kT [keV]$$= 0.85\pm0.22$~$+$~$(2.02\pm 0.28)$~$log(SA_{*}/SA_{sol})$,
with NGC 1333 giving  
 $kT [keV]$$= 1.09\pm0.26$~$+$~$(1.76\pm 0.22)$~$log(SA_{*}/SA_{sol})$.

There is also a dependence of plasma temperature on evolutionary class.  
In NGC 1333, the Class II and  Class III sources  have statistically distinguishable values of $kT$, with a 
Kolmogorov-Smirnov test giving a probability of 3.3\% that the Class II and III come from 
the same distribution (in Serpens the probability is 21\%, probably due to the small sample).   
A similar difference in the electron temperature of CTTS and WTTS was observed in the Taurus XEST 
study by \citet{tell}.   In NGC 1333, the mean value of the Class II is  $kT$ $=  2.41\pm1.19$~keV  and 
the mean value for the Class  III sources is $kT$ $= 1.12\pm0.59$~keV.  The mean value for the 3 
Class I objects is $kT$ $= 3.95\pm2.85$~keV, though the Class I objects likely have higher extinctions 
that may absorb most of the low energy X-rays.    All but one of the Class III diskless members have 
plasma temperatures $\leq 2.5$~keV, while the Class I sources all exhibit $kT$ values above 3~keV.

\subsection{X-Ray Luminosity with Isochronal Age and Mass}

It is known that $L_X$ decreases with age over the stellar lifetime and increases with stellar mass 
\citep{mica,micb}. The ages and masses of the stars in Serpens and NGC 1333 were determined by 
use of their spectral types to map them from the H-R diagram with isochrones and mass tracks taken 
from \citet{bar}, and are thus dependent on the choice of model.  The apparent isochronal ages and 
the inferred age spreads, in particular, are controversial \citep{chab,hart,pete,win08a}.     
We do not expect to see evidence for a rapid decrease in $L_X$ with age over the isochronal ages of these stars; 
other cluster studies have shown that the X-ray luminosity does not decrease rapidly until $\sim$100~Myrs \citep{mica}.
A decreasing trend in X-ray luminosity with stellar age was found for PMS stars in Orion by \citet{prefei}. 
Given their large sample, they use four mass bins from 0.1 to 2~$M_{\odot}$, while here we will 
examine the sample as a whole.  Over all four bins, they find a decreasing trend of $L_X$ with age, with 
average value: $log(L_X [erg s^{-1}])$~$\propto$~$(-0.30\pm0.13)$~$log(\tau [Myr])$.

\begin{figure}
\epsscale{1.}
\plottwo{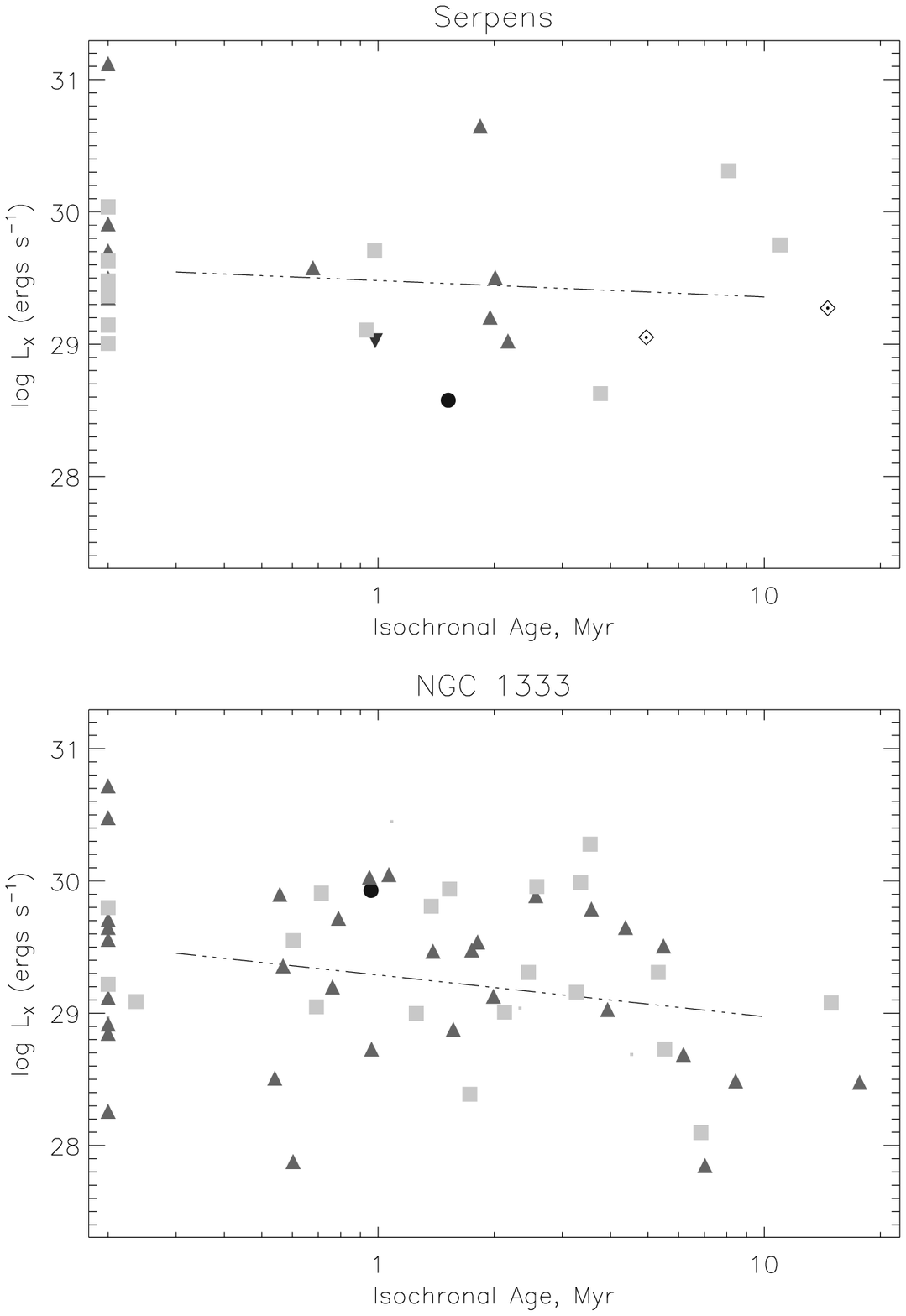}{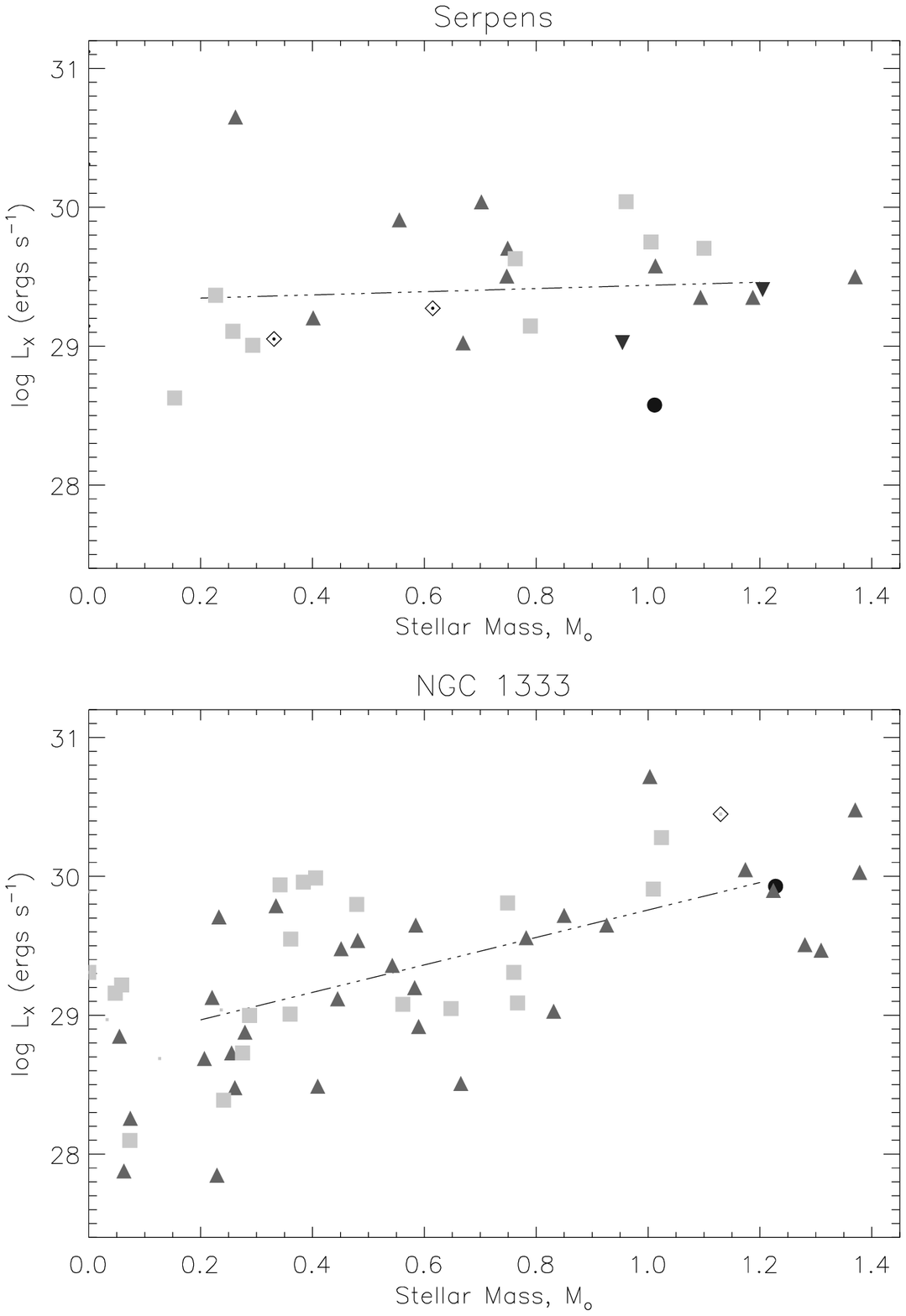}
\caption[X-ray Luminosity by Age and Mass]{ 
The graphs on the left show the stellar isochronal age calculated from \citet{bar}, plotted against the 
stellar X-ray luminosity, $L_{X}$, for Serpens (above) and NGC 1333  (below).  The NGC 1333 data 
exhibit a marginal trend towards lower luminosities in apparently older stars, the Serpens data do not 
show a significant trend.  The graphs on the right show the stellar mass calculated from \citet{bar}, 
$M_{{*}}$, plotted against $L_{X}$, for Serpens (above) and NGC 1333  (below).   
In NGC 1333, there is a weak trend towards increasing X-ray luminosity with increasing stellar mass.
Symbols indicate Class 0/I (circle), Flat Spectrum (inverted triangle), Class II (triangle), 
Transition Disk (diamond), Class III (square).  
The dot-dashed lines represent the fits to the entire YSO sample of $L_{X}$ to stellar isochronal age 
and mass:  In NGC 1333 $log(L_X [erg s^{-1}])$~$\propto$~$(-0.32\pm0.17)$~$log(\tau [Myr])$ and $L_X$ $\propto$,  
$(0.99\pm 0.16)~log(M/M_{\odot})$,  in Serpens $log(L_X [erg s^{-1}])$~$\propto$~$(-0.12\pm0.16)$~$log(\tau [Myr])$,  
$(0.12\pm 0.21)~log(M/M_{\odot})$.

}
\label{xlumagemass}
\end{figure}

Figure~\ref{xlumagemass} (left) examines the X-ray luminosities of the young stars with respect to their ages.  
A marginal trend towards lower luminosities with increasing age is observed in the NGC 1333 data, 
$log(L_X [erg s^{-1}])$~$\propto$~$(-0.32\pm0.17)$~$log(\tau [Myr])$, while no significant trend is found in the 
Serpens region. 
Since stars contract as they age, this may result from the dependence of X-ray luminosity on the surface area.     
No difference was observed in the fits of $L_X$ with isochronal age separated by evolutionary class.

\citet{pre2002} found in IC 348 that the $L_X$ $\propto$ $M^{\alpha}$ where $\alpha = 2$, whereas \citet{pre} 
found a lower $\alpha$ of 1.44 or 1.13 for stars with mass $<2~M_{\odot}$in Orion 
(depending on  whether the \citet{sie} or  \citet{pal} PMS model tracks, respectively, are used).  
Work in Taurus by \citet{tell} on the {\it XEST} survey found  $\alpha = 1.49\pm0.07~log(M/M_{\odot})$.
Figure~\ref{xlumagemass} (right) examines the effect of stellar mass on the X-ray luminosity; the masses were 
determined by \citet{win08a} using the \citet{bar} tracks.  A  trend to increasing $L_{X}$ with increasing mass 
is evident in NGC 1333, with  $\alpha = 0.99\pm 0.16$.  In Serpens  $\alpha = 0.12\pm 0.21$, we do not find 
a significant increase with mass.    The trends with mass were not found to be dependent on evolutionary class.

\subsection{Discussion of X-ray Properties}

From our analysis of the X-ray and spectral properties of the two clusters NGC 1333 and Serpens 
we identify four main relations of interest that we will discuss in this section.  

{\bf Dependence of X-ray Luminosity on Bolometric Luminosity \& Stellar Surface Area.}  
In Sec~\ref{fxlbol} we find that the $L_X$ v. $L_{bol}$ relation holds to that published in the literature 
for other young stellar clusters in Serpens and NGC 1333.  In the young stars, $log(L_{X})$ increases 
approximately linearly with $log(L_{bol})$, while for more evolved field stars the relation is shallower.   
We find that the strong dependance of $L_{X}$ on bolometric luminosity is due for the most part to an 
approximately linear dependence on the surface areas of the young stars, while there is a much weaker 
dependence with effective temperatures.   The dependence on surface area suggests that coronal X-ray 
production is saturated in the young stars.   Further, this dependence suggests that coronal emission 
(and not accretion emission) is the dominant mechanism involved in X-ray production in the YSOs.   

{\bf Dependence of X-ray Luminosity on Evolutionary Class.}    We use our IR SED classifications of Class II 
and Class III objects to compare the properties of diskless and disk-bearing YSOs and find that 
for a given surface area the Class III stars are more X-ray luminous than the Class II. 

In NGC 1333, we find some evidence that for a given surface area, the mean luminosity of the Class III YSOs is somewhat 
higher than that of the Class IIs.  The difference between the median Class II and Class III values is $\sim$0.3 dex.  
Such a difference has been noted previously in other regions and suggested explanations include: 
absorption of  X-rays due to the inclination of the circumstellar disk in Class II sources, or  
magnetic disk-braking causing slower rotation in Class IIs and thus a weaker $\alpha$-$\omega$ dynamo. 
It is also possible that the accretion column onto the surface of the Class II stars is reducing the total 
volume of the corona available to produce X-rays \citep{greg}.  This would require that $\sim$30\% (0.3~$dex$) of 
the surface area of the Class II stars was covered by accretion columns.  However, studies such as 
\citet{cal} and \citet{muz} estimate that from 1\%-10\% of the stellar surface is affected by accretion.  
This effect is therefore not large enough to fully account for the difference between the Class II and III 
stars.    

Another possible origin of the difference in Class II and III $L_X$ for a given $L_{bol}$ or $SA$ is that the 
disk-bearing YSOs are over-luminous in $L_{bol}$ due to veiling, which leads to an over estimation of their 
bolometric luminosity \citep{cie}.  In \citet{win08a} we estimate veiling of up to 0.3~$dex$ in $L_{bol}$.  However, if this 
were the case the Class II would be systematically intrinsically fainter than the Class III (by $\sim0.3$~$dex$) 
and would therefore have a greater median age ($\sim$4.5~Myr compared with the current $\sim$2.1~Myr), 
making them significantly older than the Class III ($\sim$2.4~Myr).

{\bf Dependence of X-ray Luminosity on Effective Temperature.}   In NGC 1333, we find some evidence for a 
discontinuity in the distribution of X-ray luminosities at 3800~K, the boundary between K7 and M0 
spectral classes.  The K-M boundary at 1~Myr is the approximate location of the turnover from the convection 
tracks to the radiative tracks \citep{bar}.   The discontinuity in X-ray luminosity could then be due to the transition 
between the fully convective YSOs and those developing a radiative core and convection zone.

{\bf Dependence of Plasma Temperature with Evolutionary Class, Surface Area, Effective Temperature}   
We find that the plasma temperature increases with the surface area squared, but there is no apparent 
dependence on effective temperature.   In NGC 1333, the Class III stars exhibit consistently cooler plasma 
temperatures than the Class II.  This is also seen to a lesser extent in Serpens where the Class III objects 
do not show plasma temperatures exceeding  $\sim$2.5~keV.    
If the plasma temperature is dependent on the coronal loop length, this might imply that the 
Class II stars had longer loop lengths. 
However, the higher plasma temperature and lower luminosities of Class II objects may be due  
to under-correction of disk absorption in the X-ray modelling, thus leading to hotter plasma 
temperatures (since the lower energy X-rays are absorbed) and lower $L_X$ determinations for 
the Class IIs.

{\bf Dependence of X-ray Luminosity on Isochronal Age \& Mass.}   Lastly, we find some weak dependence 
of X-ray luminosity with isochronal age and mass, qualitatively consistent with dependence on 
surface area and effective temperature, from which the values of age and mass are determined.  
The X-ray luminosity tends to decrease slightly if at all with isochronal age and to increase with mass.  
Since the isochronal age depends on surface area, the dependence of $L_X$ with age may really be the 
dependence of $L_X$ with stellar surface area.  In turn, the surface area depends in part on age, mass, and 
potentially the accretion history \citep{chab09}, thus weakening the dependence on the actual age of the source.  
The dependence of $L_X$ with stellar mass may also reflect the dependence with surface area, although the 
possible increase of $L_X$ with $T_{eff}$ may also contribute.

\section{Hydrogen Column Density and Extinction in NGC 1333 and Serpens}

\begin{figure}
\epsscale{.75}
\plotone{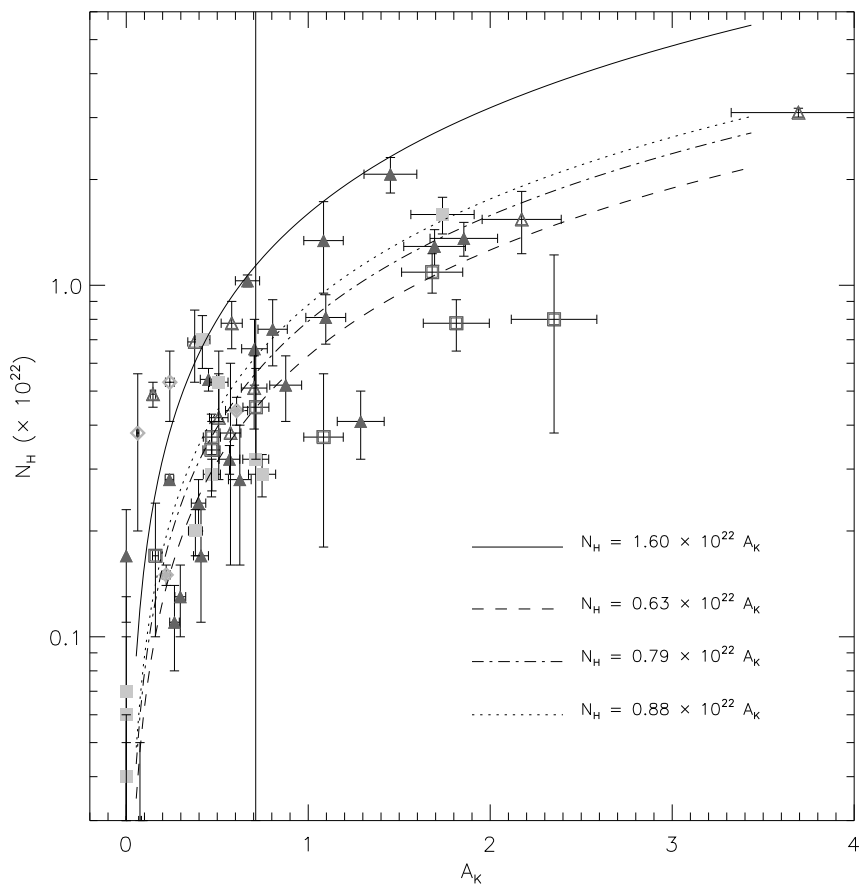}
\caption[$N_H$ - $A_K$ of Serpens \& NGC 1333]{ 
The graph plots the $N_H$ v. $A_K$ ratio for the sources from both the Serpens (open symbols) and 
NGC 1333  (filled symbols) clusters. The squares represent Class III, the triangles Class II, and the diamonds 
Transition Disk cluster stars.  Note that the Class III objects trace only the lower tracks ($N_H$ $= 0.88\times10^{22}$~$A_K$ 
for NGC 1333, $N_H$ $= 0.63\times10^{22}$~$A_K$ for Serpens, $N_H$ $= 0.79\times10^{22}$~$A_K$ for both clusters), 
while the Class II objects trace both the lower relation and the standard ISM gas-to-dust ratio of $N_H$ $= 1.6\times10^{22}$~$A_K$ \citep{vuo}. 
}
\label{figxsn}
\end{figure}

The X-ray and infrared results have also been used to calibrate the relationship between Hydrogen 
gas column density in the line of sight and the extinction measured using the near-IR bands (Fig.~\ref{figxsn}).  
Previous measurements of  this value lead  to an approximately linear fit of $1.6\times10^{21} A_{V}$ 
\citep{vuo,fei} for star forming regions, while the value for the diffuse ISM ranges from 
$1.8-2.2 \times10^{21} A_{V}$ \citep{vuo,gore}. 
In NGC 1333, \citet{pre03} noted that SVS16 had a value of $N_H$ corresponding to an $A_V \sim 10~mag$  
inconsistent with the extinction calculated in the IR of  $A_V \sim 26-28~mag$ and suggested this may be due 
to the X-ray production mechanism or a change in the canonical dust-to-gas ratio.   
\citet{winston} found the value using Class III objects in the Serpens region $N_H =$$0.63\pm0.23 \times 10^{22}$$A_K$, 
where $A_K \sim 1/10 A_V$ \citep{rieke}.  The rationale for using the Class III objects was that they would be unaffected 
by local extinction due to disks and accretion flows.  \citet{winston} proposed that this discrepant value was 
due to changes in the extinction law due to the growth of grains through the accretion of volatiles onto grain 
mantles and/or coagulation.   A limitation of this result was that the fit of the extinction law was driven by only 
three Class III objects with 1.6 $>A_K>$ 2.4.

We  calculated $A_K$ for each star using the method of \citet{gut1}, which is based on 
the reddening loci from  \citet{mey} and the extinction law of \citet{fla}.  
These values are compared to the column density of hydrogen atoms, $N_H$,  
which is calculated from the inferred  absorption of the X-ray emission for 26 objects in NGC 1333 and 
18 in Serpens. 
The results for NGC 1333 are consistent with those in Serpens by \citet{winston}.
The linear fit of the slope for seven Class III stars was  $N_H$ $= 0.89\pm0.13\times10^{22}$~$A_K$, 
while that for the Class II objects (18 in number) was $N_H$ $= 0.87\pm0.17\times10^{22}$~$A_K$.
In all cases, the ratio falls below that found for the diffuse ISM \citep{vuo}. 
In the Class II objects, the X-rays may be absorbed by the gas in an accretion flow 
(within which it would be too hot for dust grains to exist) or in a disk.  In addition, some of the near-IR 
light might be scattered by dust in the disks, potentially resulting in an underestimate of the extinction.  
This would be consistent with our analysis in Sec. 3, which shows a lower X-ray luminosity and higher X-ray 
temperature for the Class II sources compared to the Class III sources.  Since the $N_H$ v. $A_K$ relationship 
for Class II objects could be affected by the circumstellar environment, it is best determined using Class III objects.  
These diskless sources are reliably identified with the mid-IR data from {\it Spitzer}.    

The relationship of $N_H$ to $A_K$ was previously examined in the NGC 1333 region by 
\citet{get2} using an 37.5~ks {\it Chandra} observation of the region. While they find their results 
to be consistent with standard ISM values, they also note that some sources are consistent with the 
lower $N_H$ v $A_K$ slope we determined from the Serpens region.  We have compared their 
values of $N_H$ and $A_V$ to our own and found that they agree to within the 1~$\sigma$ uncertainties.  
In our work, the low $N_H$ vs. $A_K$ slope becomes apparent due to the separation of sources into Classes II 
and III and our extension to higher $A_{K}$.  
 
For the combined clusters, the slope of $N_H$ to $A_K$ is $\sim$1/3 that of the interstellar ISM.
Using the Class III objects of NGC 1333 and Serpens, we calculate that 
 $N_{H}$ $=$ $0.79\pm0.19\times10^{22}$ $A_{K}$.   
 These results are consistent with those found in the deeply embedded stars in the RCW 108 region \citep{wol2}, 
 suggesting that the slope of $N_H$ v $A_K$ is typically lower in deeply embedded regions where the extinction
 exceeds  $1~A_K$.

\section{\bf Conclusion}

In this paper we have presented {\it Chandra} X-ray observations of the NGC 1333 star forming region, and compared 
these data with the {\it Spitzer} catalogue of identified young stellar objects in the region.   
The work further presents a comparative study of the X-ray and fundamental stellar (bolometric luminosity, effective 
temperature, age, mass) properties of the NGC 1333 and Serpens clusters determined with far-red and near-IR spectra. 

\begin{itemize}

\item  In total, we identify 95 NGC 1333 cluster members in X-rays. Five of these detections were Class I, eight Flat Spectrum, 
thirty-nine Class II, and two Transition Disk objects.  We identify 41 Class III members of the NGC 1333 cluster 
through their elevated X-ray emission, 19 of which are new to this study.  This brings the total number of detected cluster 
members to 178.   

\item  We found the fraction of each class detected in X-rays; 23\% of Class I, 53\% of Flat Spectrum, 52\% of Class II, 
50\% of Transition Disks.   If the detection rate of Class II and Class III sources are the same, the total Class III population 
of the cluster is approximately  84 members (implying 221 members in total); however, if the Class III sources are more 
X-ray luminous than the Class II sources then the number of Class III sources will be lower.  None of the known Class 0 
protostars were detected by {\it Chandra}.  

\item  Using the XLF of Serpens, a new distance to the cluster is estimated at $360^{+22}_{-13}$~pc. This distance is similar 
to previous literature distances to the cluster, and is found to be compatible with the HRD of the region, giving a 
median age of 1~Myr for the star forming region.  This new distance may also be applied to the recently discovered 
Serpens South cluster.   

\item  The $L_X \propto L_{bol}$ relation found in other star forming regions was confirmed for the two clusters.  This was 
found to be predominantly a dependence of X-ray luminosity on the stellar surface area, with 
some weak dependence on the effective temperature.   There is some evidence for a jump in $L_X$ with $T_{eff}$ near 
the M0-K7 transition, which may be due to the turnover from purely convective YSOs to one with a developing radiative core.   

\item  A marginal dependence of X-ray luminosity with evolutionary class suggests that the diskless Class III sources 
in NGC 1333 are, for a given stellar surface area,  more luminous than their disk-bearing Class II counterparts. 
In Serpens the two classes show similar luminosities, although this may be due to the smaller sample size and the 
larger scatter.   

\item  The temperature of the X-ray emitting plasma was also examined as a function of effective temperature 
and surface area, with increasing trends observed with surface area.   The Class III sources are found to have lower 
values of $kT$ than the Class II in each case, with K-S probabilities of them arising from the same distribution of 21\% 
(Serpens) and 3.3\% (NGC 1333).  This may be a result of underestimation of the absorption due to circumstellar disks.   
 
\item  The variation of X-ray luminosity with age was considered, indicating a marginal trend in NGC 1333  
to higher levels of X-ray flux at younger ages.  This may reflect the dependence of isochronal age on stellar surface area.   
There is also a  trend in $F_X$ with mass, with more massive YSOs showing higher X-ray fluxes.   This is consistent with 
previous studies such as the {\it COUP} in Orion.   

\item  In the case of NGC 1333, the ratio of $N_H$ to $A_K$ is lower than expected,  $N_H$ $= 0.89\pm0.13\times10^{22}$~$A_K$,  
a similar value to that found in Serpens \citep{winston}.   The median value  for both clusters is $N_H$ $= 0.79\pm0.19\times10^{22}$~$A_K$.
This may be the result of grain growth through coagulation of the accretion of grain mantles as discussed in \citet{winston}.  

\end{itemize}

This work is based on observations made with the {\it Chandra} Telescope, under NASA contract NAS8-03060. 
This work is based on observations made with the {\it Spitzer} Space Telescope (PID 6, PID 174), 
which is operated by the Jet Propulsion Laboratory, California Institute of Technology under NASA 
contract 1407. Support for this work was provided by NASA through contract 1256790 issued by 
JPL/Caltech. Support for the IRAC instrument was provided by NASA through contract 960541 issued 
by JPL.
This work is based on observations taken with the {\it Hectospec} instrument on the MMT, a joint 
venture of the Smithsonian Institute and the University of Arizona.  
This publication makes use of data products from the Two Micron All Sky Survey, which is a 
joint project of the University of Massachusetts and the Infrared Processing and Analysis 
Center/California Institute of Technology, funded by the National Aeronautics and Space 
Administration and the National Science Foundation.
This research has made use of the NASA/IPAC Infrared Science Archive, which is operated by 
the Jet Propulsion Laboratory, California Institute of Technology, under contract with the 
National Aeronautics and Space Administration.

\clearpage

\center
\begin{deluxetable}{cccccccccc}
\tablecolumns{10}
\tabletypesize{\scriptsize}
\tablewidth{0pt}
\tablecaption{NGC 1333: {\it Chandra} X-ray Source Photometry  \protect\label{tablenp}}
\tablehead{\colhead{{\it Chandra}}  & \colhead{RA}   & \colhead{Dec} & \colhead{Raw}  &   \colhead{Net\,\tablenotemark{a}}  &   \colhead{$N_H$}   &   \colhead{kT}    &   \colhead{Abs. Flux}  &   \colhead{Tot. UnAbs. Flux}  &   \colhead{$\chi^{2}$}  \\
\colhead{ID}  & \colhead{(2000)}   & \colhead{(2000)}   & \colhead{CNT}  &   \colhead{CNT}  &   \colhead{$10^{22}cm^{-2}$}  &   \colhead{keV}  &   \multicolumn{2}{c}{$log([erg cm^{-2} s^{-1}])$}   &   \colhead{} 
}

\startdata

1  &        3:29:03.03  &       31:15:17.09  &          47  &      40.2  &      4.18$\pm$\nodata  &  12.09$\pm$\nodata  &     -14.06  &     -13.72  &      1.01   \\
2  &        3:28:52.16  &       31:15:47.75  &          36  &      29.5  &     0.10$\pm$0.14  &     0.96$\pm$0.09  &     -14.73  &     -14.58  &     0.80   \\
3  &        3:28:59.32  &       31:15:49.56  &        1552  &      1544  &      1.46$\pm$0.08  &      2.39$\pm$0.16  &     -12.79  &     -12.36  &      2.23   \\
4  &        3:29:02.91  &       31:16:01.63  &         273  &      268   &      0.00$\pm$0.00  &      1.00$\pm$0.04  &     -13.76  &     -13.76  &      1.00   \\
5  &        3:29:02.18  &       31:16:12.15  &          71  &      65.7  &    0.02$\pm$0.07  &     0.84$\pm$0.04  &     -14.40  &     -14.36  &     0.72   \\
6  &        3:29:21.15  &       31:16:17.47  &          62  &      50.2  &      18.07$\pm$\nodata  &       \nodata  &     -13.76  &     -13.21  &      1.24   \\
7  &        3:29:03.50  &       31:16:18.34  &          52  &      47.4  &      2.94$\pm$0.77  &      2.63$\pm$1.14  &     -14.18  &     -13.66  &     0.77   \\
8  &        3:29:16.72  &       31:16:19.45  &         408  &      400   &      0.00$\pm$0.02  &      1.08$\pm$0.03  &     -13.58  &     -13.58  &      1.32   \\
\enddata

\tablenotetext{a}{Net Counts: background subtracted and aperture corrected from raw counts.  }
\tablenotetext{b}{ Abs. Flux: Absorbed flux. Tot. UnAbs. Flux: Total unabsorbed flux.  }

\end{deluxetable}
\endcenter

\clearpage

\center
\begin{deluxetable}{cccccccccccccc}
\tablecolumns{14}
\rotate
\tabletypesize{\scriptsize}
 \setlength{\tabcolsep}{0.06in}
\tablewidth{0pt}
\tablecaption{NGC 1333: Coordinates \& IR Photometry of X-ray sources  \protect\label{tablenpi}}
\tablehead{\colhead{{\it Ch.}}  & \colhead{RA}   & \colhead{Dec}  & \colhead{RA}   & \colhead{Dec}  & \colhead{Sep.\,\tablenotemark{b}}  &   \colhead{$J$}  &   \colhead{$H$}  &   \colhead{$K$}  &   \colhead{$3.6{\mu}m$}  &   \colhead{$4.5{\mu}m$}  &   \colhead{$5.8{\mu}m$}  &   \colhead{$8.0{\mu}m$} &   \colhead{$24{\mu}m$}  \\
\colhead{ID}   &   \multicolumn{2}{c}{X-ray (2000)}   &   \multicolumn{2}{c}{Infrared (2000)\,\tablenotemark{a}}  & \colhead{$''$}  &  \colhead{$mag$} &  \colhead{$mag$}  &  \colhead{$mag$}  &  \colhead{$mag$}  &  \colhead{$mag$}  &  \colhead{$mag$}  &  \colhead{$mag$}  &  \colhead{$mag$}  }  

\startdata

 1   &     3:29:03.03  &     31:15:17.09   &    3:29:02.97  &     31:15:16.73   &         0.364   &    \nodata  &  \nodata &  \nodata & 13.356$\pm$0.112   & 11.9689$\pm$0.069  & 11.524$\pm$0.025 &  11.475$\pm$0.052 &  \nodata   \\
 2   &     3:28:52.16   &    31:15:47.75   &    3:28:52.13   &    31:15:47.18   &   	0.157   &   13.161$\pm$0.021   &   12.469$\pm$0.026   &   12.029$\pm$0.022   &   11.356$\pm$0.005   &   11.112$\pm$0.006   &   10.804$\pm$0.013   &    10.001$\pm$0.001   &   7.8350$\pm$0.033   \\ 
 3   &     3:28:59.32   &    31:15:49.55   &    3:28:59.30   &    31:15:48.52   &   	0.198   &   16.490$\pm$ 0.16   &   12.528$\pm$0.027   &   10.437$\pm$0.019   &   8.5941$\pm$0.002   &   8.0706$\pm$0.003   &   7.4837$\pm$0.003   &    6.4121$\pm$0.002   &   2.4270$\pm$0.032   \\ 
 4   &     3:29:02.91   &    31:16:01.64   &    3:29:02.89   &    31:16:01.03   &   	0.268   &   12.838$\pm$0.019   &   12.177$\pm$0.033   &   11.943$\pm$0.028   &   11.541$\pm$0.043   &   \nodata         &      10.843$\pm$0.115   &    \nodata         &      \nodata            \\ 
 5   &     3:29:02.18   &    31:16:12.15   &    3:29:02.16   &    31:16:11.43   &   	0.143   &   14.526$\pm$0.030   &   13.865$\pm$0.040   &   13.567$\pm$0.034   &   13.279$\pm$0.034   &   12.692$\pm$0.057   &   \nodata         &       \nodata         &      \nodata            \\
 8   &     3:29:16.72   &    31:16:19.45   &    3:29:16.67   &    31:16:18.29   &   	0.113   &   11.443$\pm$0.021   &   10.748$\pm$0.028   &   10.431$\pm$0.020   &   10.166$\pm$0.003   &   10.159$\pm$0.004   &   10.152$\pm$0.009   &    10.124$\pm$0.011   &   \nodata            \\
 9   &     3:28:56.99   &    31:16:22.97   &    3:28:56.94   &    31:16:22.28   &   	0.145   &   13.760$\pm$0.027   &   11.923$\pm$0.024   &   11.081$\pm$0.021   &   10.453$\pm$0.003   &   10.275$\pm$0.004   &   9.9456$\pm$0.008   &    8.8774$\pm$0.005   &   3.9040$\pm$0.031   \\
\enddata

\tablenotetext{a}{ The IR detections in multiple bands are matched to $<$1", with the coordinates at the shortest detected wavelength listed. }
\tablenotetext{b}{ Three sources match to infrared detections with separations $>$1":  12, 135, 188.}

\end{deluxetable}
\endcenter

\clearpage

\center
\begin{deluxetable}{cccccc}
\tablecolumns{6}
\tabletypesize{\scriptsize}
\tablewidth{0pt}
\tablecaption{NGC 1333: IR Coordinates \& {\it Chandra} ID of X-ray Detected YSOs  \protect\label{tableids}}
\tablehead{\colhead{{\it Spitzer} ID\,\tablenotemark{a}}  & \colhead{RA (2000)}   & \colhead{Dec (2000)}  & \colhead{Class\,\tablenotemark{b}} & \colhead{{\it Chandra} ID}  &   \colhead{G.ID\,\tablenotemark{c}} }
\startdata

18  &  3:28:43.26  &  31:17:33.11  &  I  &  23    &    9    \\
25  &  3:28:58.41  &  31:22:17.64  &  I  &  89     &    38   \\
30  &  3:29:04.94  &  31:20:38.59  &  FS  &  67    &   57    \\
31  &  3:29:07.76  &  31:21:57.34  &  I  &  82      &    \nodata  \\
34  &  3:29:09.08  &  31:23:05.66  &  I  &  103     &    63  \\
36  &  3:29:12.94  &  31:18:14.55  &  I  &  34     &    71   \\
45  &  3:28:43.56  &  31:17:36.53  &  FS  &  26     &    10   \\
46  &  3:28:44.08  &  31:20:52.89  &  II  &  137     &    11   \\
\enddata

\tablenotetext{a}{Identifiers from \citet{gut07}.}
\tablenotetext{b}{Evolutionary classification modified from that of \citet{gut07} based on \citet{winston}. I: Class I, FS: Flat Spectrum, II: Class II, TD: Transition Disk, III: Class III.}
\tablenotetext{c}{Identifiers from \citet{get2}.}

\end{deluxetable}
\endcenter

\clearpage

\center
\begin{deluxetable}{ccccccccccccc}
\tablecolumns{13}
\tabletypesize{\scriptsize}
\tablewidth{0pt}
\tablecaption{Serpens: Spectroscopic \& X-ray Properties of YSOs   \protect\label{tableserprop}}
\tablehead{  \colhead{{\it Sp} ID\,\tablenotemark{a}} & \colhead{SpT} & \colhead{Class\,\tablenotemark{b}} & \colhead{{\it Ch} ID}  & \colhead{log($\frac{L_{bol}}{L_{\sun}}$)}   & \colhead{$T_{eff}$}  & \colhead{log($\frac{SA}{SA_{\sun}}$)}  &   \colhead{$L_{X}$}  &  \colhead{kT}   &  \colhead{$N_{H}$}   &  \colhead{$A_{K}$\,\tablenotemark{c}}   &  \colhead{Age\,\tablenotemark{d}}   &  \colhead{Mass\,\tablenotemark{d}}  \\
\colhead{}  &  \colhead{}  &  \colhead{}  &  \colhead{}  &  \colhead{}  &  \colhead{[K]}   &  \colhead{}  &  \colhead{[$erg s^{-1}$]}  &  \colhead{[keV]}  &  \colhead{[$\times 10^{22}$]}  &  \colhead{[mag.]}  &  \colhead{[Myr]}  &  \colhead{[$M_{\sun}$]}   
 }
\startdata

      9      &  K8.0$\pm$1.5    &  0/I              &        67       &   -0.179  &    3892   &      3.22  &    28.57   &     1.30   &      0.86   &     \nodata &  1.5      &    1.01       \\     
      27    &  K7.0$\pm$3.5    &  FS         &        74       &    0.265  &    3997   &      8.09  &    29.40   &     2.01   &      0.65   &     0.94    &  $\sim$0.2  &    1.20        \\     
      33    &  M4.5$\pm$1.0   &  FS          &        \nodata  &   -0.867  &    3197 &      1.45  &    \nodata  &    \nodata  &    \nodata  &   0.11    &  1.4      &    0.23        \\     
      36    &  K6.5$\pm$2.5    &  FS          &        9        &    0.853  &    4147   &      27.1  &    \nodata  &    \nodata  &    \nodata  &   0.95    &  $\sim$0.2  &    1.13       \\     
      40    &  M0.0$\pm$1.5   &  FS          &        44       &   -0.090  &    3821   &      4.26  &    29.02    &    1.80     &    0.86     &   0.94    &  0.9     &    0.95       \\     
      53    &  M4.0$\pm$2.0   &  II             &        \nodata  &   -1.281  &    3299   &      0.49  &    \nodata  &    \nodata  &    \nodata  &   0.21    &  9.0      &    0.25       \\     
      54    &  M9.0$\pm$5.0   &  II             &        \nodata  &   -1.778  &    2400   &      0.56  &    \nodata  &    \nodata  &    \nodata  &   0.27    &  $\sim$0.2  &    0.02       \\     
      59    &  K7.5$\pm$1.5    &  II             &        77       &    0.356  &    3955   &      10.4  &    29.35    &    4.22     &    0.38     &   0.57    &  $\sim$0.2  &    1.18       \\
\enddata

\tablenotetext{a}{{\it Sp}: {\it Spitzer}, {\it Ch} {\it Chandra}.   See Tables 6 \& 4 of \citet{win08a} for coordinates and magnitudes associated with spectra.}
\tablenotetext{b}{0/I: Class 0/I, FS: flat spectrum, II: Class II, TD: Transition Disk, III: X-ray Detected Class III. The [III] are the new candidate Class IIIs with detected $Li~I$ absorption. }
\tablenotetext{c}{A value of zero indicates that the source was blueward of the CTTS locus, a null value indicates a value could not be calculated.}
\tablenotetext{d}{Isochronal Age and Mass are determined by interpolation from the \citet{bar} tracks. Age of $\sim$0.2 indicates the source is very young but on edge of isochrone grid.}

\end{deluxetable}
\endcenter

\clearpage

\center
\begin{deluxetable}{ccccccccccccc}
\tablecolumns{13}
\tabletypesize{\scriptsize}
\tablewidth{0pt}
\tablecaption{NGC 1333: Spectroscopic \& X-ray Properties of YSOs   \protect\label{tablen13prop}}
\tablehead{\colhead{{\it Sp} ID\,\tablenotemark{a}} & \colhead{SpT} & \colhead{Class\,\tablenotemark{b}} & \colhead{{\it Ch} ID\,\tablenotemark{b}}  & \colhead{log($\frac{L_{bol}}{L_{\sun}}$)}   & \colhead{$T_{eff}$}  & \colhead{log($\frac{SA}{SA_{\sun}}$)}  &   \colhead{$L_{X}$}  &  \colhead{kT}   &  \colhead{$N_{H}$}   &  \colhead{$A_{K}$\,\tablenotemark{c}}   &  \colhead{Age\,\tablenotemark{d}}   &  \colhead{Mass\,\tablenotemark{d}}  \\
\colhead{}  &  \colhead{}  &  \colhead{}  &  \colhead{}  &  \colhead{}  &  \colhead{[K]}   &  \colhead{}  &  \colhead{[$erg s^{-1}$]}  &  \colhead{[keV]}  &  \colhead{[$\times 10^{22}$]}  &  \colhead{[mag.]}  &  \colhead{[Myr]}  &  \colhead{[$M_{\sun}$]}   
 }

\startdata

18     &      K7.0$\pm$1.0   &  0/I        &     23      &    0.066     &      4089    &     4.670     &      29.93     &     3.95     &    2.78     &   \nodata     &    0.9       &     1.23      \\  
45     &      M5.0$\pm$1.0   &  II       &     26      &   -0.420     &      3125    &     4.461     &      29.71     &     3.73     &    0.35     &    0.41     &    $\sim$0.2 &     0.23      \\  
46     &      M7.5$\pm$1.0   &  II       &     137     &   -1.533     &      2829    &     0.511     &      27.88     &     1.29     &    \nodata  &    0.22     &    0.6       &     0.06      \\  
47     &      M3.0$\pm$1.5   &  II       &     13      &   -0.101     &      3415    &     6.514     &      29.65     &     1.04     &    0.11     &    0.26     &    $\sim$0.2 &     0.58      \\  
48     &      M4.0$\pm$1.5   &  II       &     110      &   -0.793     &      3270    &     1.576     &      28.88     &     2.22     &    0.89     &    0.83     &    1.6       &     0.28      \\  
49     &      M8.0$\pm$1.0   &  II       &     17      &   -0.918     &      2710    &     2.504     &      28.85     &     2.32     &    0.18     &    0.27     &    $\sim$0.2 &     0.05      \\  
50     &      M4.5$\pm$1.0   &  II       &     36      &   -0.051     &      3197.5  &     9.516     &      29.12     &     6.11     &    0.17     &    0.41     &    $\sim$0.2 &     0.44      \\  
51     &      M4.0$\pm$1.0   &  II       &     128     &   -1.250     &      3255.5  &     0.560     &      27.85   &     3.25  &    \nodata  &    0.04     &    7.0       &     0.23      \\
\enddata

\tablenotetext{a}{{\it Sp}: {\it Spitzer}, {\it Ch} {\it Chandra}.   See Table \ref{tablenpi} \& Tables 2 \& 3 of \citet{gut07} for coordinates and magnitudes associated with spectra.}
\tablenotetext{b}{0/I: Class 0/I, FS: flat spectrum, II: Class II, TD: Transition Disk, III: X-ray Detected Class III. The [III] are the new candidate Class IIIs with detected $Li~I$ absorption. }
\tablenotetext{c}{A value of zero indicates that the source was blueward of the CTTS locus, a null value indicates a value could not be calculated.}
\tablenotetext{d}{Isochronal Age and Mass are determined by interpolation from the \citet{bar} tracks. Age of $\sim$0.2 indicates the source is very young but on edge of isochrone grid.}

\end{deluxetable}
\endcenter

\clearpage

\clearpage

\clearpage

\clearpage

\clearpage

\clearpage

\clearpage

\clearpage

\clearpage

\clearpage

\clearpage

\clearpage

\clearpage

\end{document}